\documentclass[letterpaper]{article} % DO NOT CHANGE THIS
\usepackage{aaai2026}  % DO NOT CHANGE THIS
\usepackage{times}  % DO NOT CHANGE THIS
\usepackage{helvet}  % DO NOT CHANGE THIS
\usepackage{courier}  % DO NOT CHANGE THIS
\usepackage[hyphens]{url}  % DO NOT CHANGE THIS
\usepackage{graphicx} % DO NOT CHANGE THIS
\usepackage{natbib}  % DO NOT CHANGE THIS AND DO NOT ADD ANY OPTIONS TO IT
\usepackage{caption} % DO NOT CHANGE THIS AND DO NOT ADD ANY OPTIONS TO IT
\usepackage{algorithm}
\usepackage{algorithmic}
\usepackage{enumitem}

\usepackage{booktabs}
\usepackage{tabularx}
\usepackage[table]{xcolor}

\usepackage{xcolor}
\newcommand{\answerYes}[1]{\textcolor{blue}{#1}} 
 
\newcommand{\answerNA}[1]{\textcolor{gray}{#1}}

\usepackage{tcolorbox}\tcbuselibrary{skins}

\usepackage{newfloat}
\usepackage{listings}
\DeclareCaptionStyle{ruled}{labelfont=normalfont,labelsep=colon,strut=off} % DO NOT CHANGE THIS
\floatstyle{ruled}
\newfloat{listing}{tb}{lst}{}
\floatname{listing}{Listing}
\title{Assessing How Hate, Counterspeech, and Toxicity Affect Hate Group Newcomers}
\author{
Daniel Hickey\textsuperscript{\rm 1},
Daniel M.T. Fessler\textsuperscript{\rm 2, 3, 4},
Matheus Schmitz\textsuperscript{\rm 5},
Paul Smaldino\textsuperscript{\rm 6, 7, 8},
Kristina Lerman\textsuperscript{\rm 9},
Goran Muri\'c\textsuperscript{\rm 5},
Keith Burghardt\textsuperscript{\rm 10}
}
\affiliations{
    \textsuperscript{\rm 1}School of Information, University of California, Berkeley\\
 \textsuperscript{\rm 2}Department of Anthropology, University of California, Los Angeles\\
 \textsuperscript{\rm 3}Bedari Kindness Institute, University of California, Los Angeles\\
 \textsuperscript{\rm 4}Center for Behavior, Evolution, \& Culture, University of California\\
 \textsuperscript{\rm 5}USC Information Sciences Institute\\
 \textsuperscript{\rm 6}Department of Cognitive and Information Sciences, University of California, Merced\\
 \textsuperscript{\rm 7}Santa Fe Institute\\
 \textsuperscript{\rm 8}Complexity Science Hub\\
 \textsuperscript{\rm 9}Luddy School of Informatics, Computing and Engineering, Indiana University\\
 \textsuperscript{\rm 10}School of Data Science, University of North Carolina at Charlotte\\
dan\_hickey@berkeley.edu, keithburg@charlotte.edu
}

\usepackage{bibentry}
\begin{document}

\maketitle

\begin{abstract}
Counterspeech has gained attention as a strategy to reduce hate speech on social media. Although previous studies suggest that counterspeech can reduce hate speech, little is known about its effects on participation in online hate communities. Relatedly, we lack an understanding about the degree of hostility in counterspeech. Hostile counterspeech may increase online conflict, potentially hardening the positions of hate adherents, and further eroding online environments. Here, we analyzed the effect of counterspeech on 16,513 newcomers across 104 hate subreddits (forums within Reddit.com). We devised an LLM-based counterspeech detection approach that outperforms specialized models trained on existing datasets, then examined the presence, and effects of, hostility. While counterspeech comments are less toxic than hate speech comments, they are almost twice as toxic as other discourse within hate subreddits. We then evaluated the effect of counterspeech on newcomer engagement in hate subreddits. We found that newcomers using hate speech who receive counterspeech are less likely to continue posting within these hate subreddits, rather than becoming galvanized. We speculate that, instead of constituting ardent hate adherents, readily-dissuaded newcomers may merely be toying with beliefs that are proscribed in other contexts. Although we found no association between the toxicity of counterspeech and its effects on user retention, consistent with prior research regarding the harmful effects of toxic speech, we found that toxic counterspeech increases the probability of continued hostility from hate users within the same discussion.
\end{abstract}

\begin{links}
\link{Code and data: }{https://github.com/dan-hickey1/toxic-counterspeech}
\end{links}

\section{Introduction}

\begin{figure}[h]
    \centering
    \includegraphics[width=0.9\columnwidth]{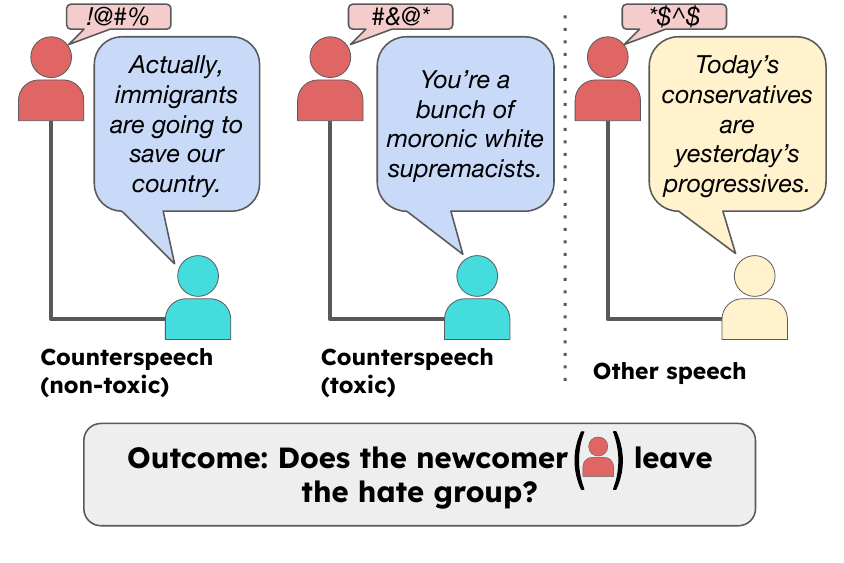}
    \caption{Assessing the impact of counterspeech on new hate group users. We compared new hate group users that receive different types of replies, such as counterspeech replies or other speech (including hate speech from other hate group users). We evaluated the association between these types of replies, as well as their toxicity levels, on newcomer retention in hate groups.}
    \label{fig:counterspeech_examples}
\end{figure}

Online hate is pervasive \cite{hawdon2024measuring}, contagious \cite{mathew2020hate}, and presents a threat to the well-being of marginalized groups on the internet \cite{gelber2016evidencing}.  Online hate speech activity has been associated with offline hate crimes \cite{chan2016internet, lupu2023offline}. Importantly, group membership plays a crucial role in the facilitation of online hate and radicalization \cite{doosje2016terrorism}. In dedicated hate groups, users can receive social approval for spreading hate speech \cite{walther2024effects}, and members can more easily organize to harass members of marginalized groups \cite{marwick2021morally}. 

Given the harms posed by online hate groups, a large body of research has focused on understanding whether various content moderation policies will reduce levels of hate in online environments. Although many studies show that content moderation strategies are effective in reducing overall hate speech levels within a single platform \cite{chandrasekharan2017you, cima2024great}, these strategies can have unintended consequences that are not yet well understood, such as mass user migrations to alt-tech platforms \cite{russo2023spillover}. Furthermore, some criticize the practice of moderating hate speech as a dangerous restriction of free speech \cite{kozyreva2023resolving}. Even if a majority of the public agrees with certain moderation policies, a platform's leadership may not be interested in implementing those policies. Consider Elon Musk's takeover of Twitter (now X), where, in the name of free speech, Musk stated that he would loosen content moderation -- following which hate speech on the platform increased by 50\% \cite{hickey2025x}.

Due to the limitations of using moderation policies to combat hate speech, counterspeech (speech that opposes hate speech) has emerged as an appealing option for reducing online hate. In addition to avoiding controversies regarding restricting speech, unlike moderation practices, counterspeech can be exercised by individual users independent of the position taken by a given social media platform. Additionally, a personalized and well-reasoned argument could be more powerful than moderation in persuading users to reject a hateful ideology. Emerging empirical evidence confirms the promise of counterspeech: both field experiments and observational studies have shown that when authors of posts containing hate speech on X receive replies containing counterspeech, they become less likely to produce hate speech in the future \cite{hangartner2021empathy, he2021racism, gennaro2025counterspeech}. However, given the variety of contexts, outcomes, and methodological approaches that are used to study counterspeech, more work must be done to achieve consensus on its effects on online hate. In particular, prior work has focused primarily on changes in hate speech production rather than downstream behavioral outcomes such as user retention, and has largely examined counterspeech in mainstream social media contexts rather than within dedicated hate communities \cite{hangartner2021empathy, he2021racism, gennaro2025counterspeech}. Moreover, existing studies rarely isolate the role of counterspeech toxicity itself, leaving open the question of whether observed benefits of counterspeech come at the cost of increased hostility or disengagement.

Previously, field experiments have been deployed to test the effects of different counterspeech strategies, such as perspective-taking, humor, and empathy-based counterspeech \cite{hangartner2021empathy, gennaro2025counterspeech}. Here, we examine a potentially problematic counterspeech strategy that has often been overlooked in studies that suggest that counterspeech is effective: responding with hostile or aggressive speech. Scholars have previously warned against this form of counterspeech, since toxic responses can attract even more toxicity within a discussion \cite{yu2024hate, saveski2021structure}. Indeed, prior research suggests that, in hate groups, toxic counterspeech can result in aggressive ``flame wars'' \cite{mann2003evolution, baider2023accountability}, indicating that, rather than effectuating a change in beliefs, venomous criticism may lead to a hardening of the target's position. This has important implications for prior studies of counterspeech. With regard to observational studies \cite{he2021racism}, if the positive effects of counterspeech are explained by the toxicity of counterspeech, the benefits of reducing hate speech may be outweighed by negative outcomes of the counterspeech's hostility that are unobserved in such studies. With regard to field experiments \cite{hangartner2021empathy, gennaro2025counterspeech}, while these studies can be useful for isolating and explaining the effects of specific counterspeech strategies, counterspeech ``in the wild'' may be far more toxic than the counterspeech tested by investigators, such that there may be substantial differences between the counterspeech that users employ and that which experiments show to be effective. 

In examining the potential effects of counterspeech on those who post hate speech in hate forums, it is useful to distinguish between users who have a long history of posting hate content, and users who are posting such content for the first time in the given location. For multiple reasons, it is plausible that longtime participants in a hate forum will be less affected by counterspeech than will newcomers. First, longtime users have a greater likelihood of having previously encountered counterspeech, whether online or offline, and thus may be more inured to its effects. Second, unlike newcomers, longtime users will have established an identity in the given forum, and thus would face greater reputational costs among their fellow hate proponents were they to be dissuaded by counterspeech from continuing to promulgate the given ideology. Lastly, as evidenced by their track records, longtime users may hold the given hate beliefs more deeply than do newcomers. Taken together, these possibilities suggest that the clearest signal of any effect of counterspeech on the posting behavior of users of hate speech will be found among newcomers to hate forums.  

Combining the above considerations, we ask the following research questions:

\begin{enumerate}[leftmargin=*, labelwidth=3em, labelsep=1em, align=left]
    \item[\textbf{RQ1}] How do toxicity levels of counterspeech found on social media compare to other forms of speech?
    \item[\textbf{RQ2}] What is the effect of (potentially toxic) counterspeech on newcomers to hate communities?
\end{enumerate}

As an initial contribution aimed at answering these questions, we conducted a causal model-based study to analyze the effect of interactions in a set of 49,539 newcomers across 104 hate subreddits and 241,940 users in 104 matched non-hate subreddits. We then employed an LLM-based prompting method to identify counterspeech and hate speech within the hate subreddits, and found that our method outperforms previous attempts to detect counterspeech on Reddit. Using the Perspective API \cite{Jigsaw2017}, we explored the nature of counterspeech within our studied subreddits. We found that, although the average toxicity of counterspeech is lower than that of hate speech, counterspeech is 88\% more toxic than posts within hate subreddits that do not employ hate speech, and over \emph{330\%} more toxic than posts in non-hate subreddits. We then studied how interactions on hate and non-hate subreddits affected the probability that a user continues to be an active member, focusing on counterspeech replies to newcomers who use hate speech in their first post. We used replies to a user's first post as a treatment and matched similar triplets of users: those who receive no reply, those who receive a reply containing counterspeech, and those who receive a reply without counterspeech. Using a mixed-effect logistic regression model, we estimated the effects of receiving replies containing (toxic or non-toxic) counterspeech on user retention, as shown in Fig.~\ref{fig:counterspeech_examples}. In hate subreddits, we found that new users who receive counterspeech replies are less likely to continue participating in the subreddit than those who do not, while replies containing hate speech make them more likely to engage. While the toxicity of counterspeech does not seem to have an effect on newcomers' further engagement in other discussion threads in their respective hate subreddits, toxic counterspeech elicits more toxic responses from hate users. Overall, these results have significant implications for counterspeech research and practice moving forward. First, the inclination of those who oppose hate speech to respond to such speech with toxicity suggests that substantial work is needed to empower counterspeakers to use more constructive discourse. Second, the dissuasive effect of counterspeech on newcomers raises the possibility that such users are not committed to hate ideologies, hence they may constitute a productive focus for interventions designed to reduce hate. 

%We release our code and data in the following repository: \url{https://github.com/dan-hickey1/toxic-counterspeech}.

\section{Related Work}

%\subsection{Online Hate Groups}
%Online hate groups have remained a pervasive problem online for decades, due to the ability of the internet to bring together members of fringe ideologies irrespective of their location. Hate groups have existed on their own dedicated platforms or websites, such as Stormfront \cite{tornberg2024inside}, as well as more mainstream platforms, such as Facebook, Reddit, and Twitter \cite{hickey2025peripatetic, phadke2020many}. Various ideologies of hate groups have existed on Reddit, including trans-exclusionary radical feminists \cite{lu2022subtle}, the manosphere \cite{ribeiro2021evolution}, and the alt-right \cite{mamie2021anti}. In the past decade, several ``alt-tech'' platforms, such as Voat, Parler, Gab, and Truth Social have emerged in response to increasingly strict moderation policies on mainstream platforms \cite{newell2016user, buntain2023cross}.

%Here, we focus on hate groups on Reddit. While Reddit is a mainstream platform, it has had an extensive history of both hosting and banning a variety of hate groups \cite{cima2024great}. Reddit's structure of hosting many subcommunities of different topics is conducive to our study design, allowing us to explore group membership across a variety of hate ideologies.

\subsection{Moderation of Hate Speech}
Although some platforms may claim to be havens for free speech, ``all platforms moderate'' \cite{gillespie2018custodians}, as the filtering, ranking, and recommendation of certain forms of speech is fundamental to user experience on social media. Despite this, platforms have exhibited varying approaches to moderating hate speech over the years, many of which have been investigated by a growing body of empirical research. In the context of hate communities, prior research has investigated community bans, which have been shown to reduce hate speech on Reddit \cite{chandrasekharan2017you}, or ``quarantines,'' (subreddit-level interventions that reduce their visibility), which reduce new user activity \cite{chandrasekharan2022quarantined, shen2022tale}. Other work highlights the unintended impacts of moderation. For example, some users will migrate platforms after community bans, and it is not well understood whether communities become more extreme after migrating platforms \cite{horta2021platform, horta2023deplatforming, russo2023understanding, johnson2019hidden, mekacher2023systemic, buntain2023cross}. Even within the same platform, certain subsets of hate community members become more toxic after community bans \cite{russo2023spillover, cima2024great}. Furthermore, \citet{trujillo2022make} show that low-quality link sharing increased after Reddit quarantined and banned r/The\_Donald. The effects of moderation also vary by platform, with moderation policies implemented on alt-tech platforms having mixed results \cite{erickson2025content, kumarswamy2025causal}.

%Because of the controversial nature of content moderation and its sometimes unsatisfying outcomes, there has been a significant focus on alternative approaches to reducing hate speech and conflict in online environments. One commonly explored idea is that of ``proactive content moderation,'' which focuses on applying interventions, such as reminders of community norms \cite{matias2019preventing}, when it is recognized that harmful actions may occur in the future. Prior work on forecasting subreddit bans \cite{habib2022proactive}, conversation incivility \cite{yu2025measuring, yu2024hate, chang2019trouble, zhang2018conversations}, or participation in hate groups \cite{hickey2025peripatetic, lahnala2025unifying} contributes to this paradigm.

\subsection{Counterspeech}
%Common lines of inquiry in counterspeech studies include testing the effectiveness of counterspeech, detecting counterspeech, generating counterspeech, and developing AI systems that enable users to write effective counterspeech. 

Studies that examine the efficacy of counterspeech can be split into two main categories: experiments that expose social media users to counterspeech messages designed by researchers \cite{hangartner2021empathy, gennaro2025counterspeech, bar2024generative}, and observational studies that use historical social media data to identify counterspeech and measure its association with various outcomes \cite{he2021racism, yu2023fine}. Both methods have been limited in the contexts in which they are applied, with most studies of counterspeech focusing exclusively on X, and the counterspeech studied primarily countering racist content. Most studies examine the hate speech levels of users as the primary outcome in examining the efficacy of counterspeech, although other studies assess responses to counterspeech \cite{yu2023fine, song2025echoes}. Notably, while studies of counterspeech show that exposure to such speech reduces future hate speech, studies of cross-partisan interactions show that they either have no effect, or have polarizing effects \cite{bail2018exposure, xia2025integrated, suhay2018polarizing}. While partisanship is not the same as hate speech, those results suggest that arguments, including counterspeech, may not result in a change of belief.

Because human-generated counterspeech can be difficult to scale, there has been a significant focus on counterspeech generation \cite{dinkar2025can}. However, there still remain key differences between machine-generated counterspeech and human-generated counterspeech \cite{alyahya2025hatred, song2024assessing}, with some studies finding machine-generated counterspeech to be less persuasive than human-generated counterspeech \cite{mun2023beyond, bar2024generative}. Much work also exists on developing technology that assists humans in writing productive counterspeech \cite{ding2024counterquill, ping2025perceiving}.

In multiple ways, we build on existing studies that seek to detect counterspeech and test its effectiveness. First, we develop an LLM-based classifier for detecting counterspeech across subreddits that differ in the identities which they target (some subreddits are misogynistic, some are racist, some are transphobic, etc.). Then, we apply causal inference methods to measure how counterspeech affects hate group dynamics, investigating whether new users stop posting in online hate communities after being exposed to counterspeech. We further add to the current research by addressing whether highly prevalent toxic counterspeech is an effective strategy.

\section{Methods}

To understand the effects of counterspeech on retention of new users in hate groups, we employed a dataset collected by Hickey et al. \citeyear{hickey2025peripatetic}, a comprehensive set of 168 hate subreddits. We sourced all of our Reddit data from Pushshift \cite{baumgartner2020pushshift}. While all subreddits in the dataset contain at least 1,000 users, the relative rarity of counterspeech means that many subreddits may not contain a sufficient amount of counterspeech to be included in our analyses. After filtering out subreddits with 10 or fewer hate speech/counterspeech pairs, we were left with 104 subreddits, representing a total of 16,513 matched triplets. Our results are robust when not filtering out any subreddits, or when using more aggressive filtering (e.g., filtering out subreddits with fewer than 100 counterspeakers, see Appendix Figure~\ref{fig:robustness_sample_size}).

% As noted above, if counterspeech has an effect, it is more likely to be evident among newcomers to hate forums, hence 
Our methodological framework can be summarized as follows (see method overview in Appendix Figure~\ref{fig:methods_schematic}): First, we identified new users in each subreddit, i.e. those who employ hate speech and comment on that subreddit for the first time. Since we are interested in measuring the effect of receiving a reply consisting of counterspeech, we next compared three categories of newcomers: those who received no replies to their comment, those who received a reply that contained counterspeech, and those who received a reply that did not contain counterspeech. We then applied causal inference matching methods described below to eliminate the effects of potential confounding variables. Finally, we used a mixed effects logistic regression model to estimate the effect of counterspeech, factoring the toxicity of counterspeech into the model estimate.

\subsection{Counterspeech Detection}

Although significant research has been conducted on counterspeech detection, the models and datasets used for the task come from a variety of domains and often only include specific kinds of counterspeech (e.g. \cite{he2021racism, mathew2019thou}). Meanwhile, LLMs have become increasingly available and have been shown to generalize well across a variety of tasks \cite{ziems2024can}. We therefore experimented with using LLMs for counterspeech detection.

We employed two methods for using LLMs for counterspeech detection: the first is a model distillation method, in which a LLM generates labels that are used to fine-tune a smaller language model. For this approach, we used the \texttt{gpt-4o-2024-08-06} model from OpenAI to label 5.5k comments from the hate subreddits to use as a training dataset. We prompted the LLM to label pairs of newcomer comments and their replies as either counterspeech, hate speech, or other speech. In our prompt, we provided the comment or submission the newcomer replied to as context, as well as the first reply to the newcomer if they received one. The complete prompt can be found in the appendix (Fig~\ref{fig:prompt}). To train a counterspeech detection model on the GPT-4o–labeled data, we followed the architecture described in \citet{yu2022hate} and fine-tuned RoBERTa \cite{liu2019roberta} with a two-layer feed-forward classification head. The classification head consists of a hidden layer with 768 units and a tanh activation, followed by an output layer with three units and a softmax activation corresponding to the three class labels. We used the concatenation of each comment to its parent (with the [SEP] token separating the comments) as input. In addition to training a model on GPT-4o labels, we benchmarked our performance against models trained on datasets obtained from prior research \cite{yu2022hate, yu2023fine}. We used an 80/20 training/validation split for each dataset. For all models, we used the hyperparameters given in \citet{yu2022hate} to ensure consistency. We used the AdamW optimizer, a dropout of 0.5, batch size of 16, and a learning rate of 1e-5 to train the model. We trained each model for 10 epochs, keeping the model with the lowest validation loss for final evaluation. 

Our second method of employing LLMs to detect counterspeech involved using an open-weight model to apply labels across our entire corpus of text. While closed models such as GPT-4o are powerful, it is prohibitively expensive to label millions of comments with them. For this reason, we chose to use Llama 3.3 70B, as it is small enough for us to feasibly label millions of comments, but large enough to be sufficiently powerful. We prioritized high precision over recall in model training, reflecting the relative rarity of counterspeech in the data. In this setting, a conservative classifier minimizes false positives, and, even with lower recall, the majority of instances predicted as non-counterspeech are expected to be correctly labeled. In other words, false positives are more costly than false negatives in our downstream analysis. For this reason, we needed to be able to differentiate between comments in which the model has high confidence and those that are borderline predictions. To achieve this, we prompted the model five times for each comment, using the number of times the model responded with a given label as the model's confidence for that label. In this setup, the temperature of the language model is a key factor in determining its usefulness. If the temperature is too low, the model will create similar answers every time, reducing the utility of prompting the model multiple times. If the temperature is too high, answers may be too inconsistent and lack coherence. Using our validation data (described below), we evaluated the precision of LLaMa 3.3 70B across different temperature values ranging from 0.1 to 2.0 in increments of 0.1. We also compared a model fine-tuned on our GPT-4o labels using Q-LoRA \cite{dettmers2023qlora} to the base LLaMa model, finding that the fine-tuned version with a temperature of 1.2 reached the highest precision value. We fine-tuned the model for 3 epochs using an $r$ value of 64, a dropout of 0.05, and an $\alpha$ value of 16, informed by the experiments conducted by Dettmers et al. \citeyear{dettmers2023qlora}. We tested $r$ values of 32 and 128, as well as dropout values of 0 and 0.1, finding the results to be relatively robust with respect to the values. For both fine-tuning and inference, we used four NVIDIA RTX A6000 GPUs on a remote computing cluster. It took approximately 12 hours to fine-tune the model.

Given that we used a variety of models trained outside of our intended domain, we needed to ensure the validity of the counterspeech and hate speech detection within the studied subreddits. We achieved this by manually annotating a validation set of 336 pairs of newcomer and reply comments (two pairs from each subreddit in the dataset), coding whether the pairs contain counterspeech, hate speech, or other speech. Three annotators independently labeled each comment, obtaining a Fleiss Kappa score of 0.52, indicating moderate agreement, consistent with much of the prior literature on hate speech detection \cite{vidgen-etal-2021-introducing}. We used the majority vote among annotators to decide on the final label for each comment. Since we are primarily interested in a counterspeech model that results in high precision estimates, and precision varies by threshold, we chose the threshold for each model that maximizes precision while ensuring that at least ten comments total are predicted as counterspeech/hate speech. In addition to reporting the precision of each model, we also report ROC-AUC values as a threshold-agnostic metric to summarize overall performance.

Given the rarity of counterspeech (approximately 7\% in our annotated sample), our initial validation dataset was too small to adequately assess precision. Additionally, since we chose our temperature values and models based on their performance on said validation dataset, we needed to ensure that we were not simply overfitting to this dataset. For this reason, we also annotated sets of 100 comments that were labeled by our final model as counterspeech at three different confidence thresholds (100 comments for each threshold) to ensure the precision of our model was sufficiently high. 

We release our data used for training/validation of the counterspeech and hate speech detection models in our GitHub repository. We adhere to FAIR principles \cite{fair} by documenting the data and making them widely available on a public platform, as well as sharing them in a widely used format (CSV).

\subsection{Matching}

\subsubsection{User matching}

To reduce the influence of confounders on the estimate of counterspeech effectiveness, we employed causal matching methods. Although most studies using matching methods have a binary treatment variable and match pairs of users together, our study design involves a categorical treatment: some newcomers receive a reply containing counterspeech, others receive a different reply, and others receive no reply at all. Matching using a categorical treatment presents a unique challenge, as optimally matching more than two groups is NP-hard. We used the method for matching categorical treatment groups outlined in \citet{nattino2021triplet}. In brief, the method starts with an arbitrary pair of treatment groups and optimally matches them, then matches units from the third treatment group that minimize the sum of the pairwise distances between units in each triplet. The triplets are then iteratively checked to see if the distance can be further minimized. Importantly, this process will not optimally match triplets, so the treatment group chosen as the initial pair could affect the final matching. For this reason, we performed the matching three times for each subreddit, starting with a different combination of two treatment groups to match each time. We chose the set of matched triplets where the absolute standardized mean difference (SMD) was below the recommended threshold of 0.1 for the greatest number of covariates \cite{austin2011introduction}.

Before matching newcomers in the hate subreddits, we filtered the newcomers in our sample to only those who were detected to be using hate speech. We performed this step for several reasons, the primary one being that explicit hate speech is much more likely to elicit counterspeech responses than are other types of speech (in our sample, comments detected as hate speech are 13.5 times more likely to receive a counterspeech reply). Furthermore, we found that our counterspeech detection model was less precise when the detected counterspeech was not a response to a comment detected as hate speech (in preliminary annotations we observed a 15 percentage point drop in precision). Finally, matching triplets of users is computationally intensive, to the point where we could not feasibly match all users within two weeks of compute time. Even in subreddits where hate speech is prevalent, hate speech still represents a minority of all comments due to the number of comments that are either missing sufficient context to be appropriately classified or discussing related topics without using explicit hate speech. Therefore, in addition to reducing the level of noise in our analysis, focusing on newcomers whom we are confident used hate speech is also more efficient.

We matched users based on thirteen potential confounders: the semantic similarity of the newcomer's comment (computed using MPNet text embedding model \cite{song2020mpnet}); the nest level of the comment (with 1 indicating a top-level comment, 2 indicating a direct reply, 3 indicating a reply to that reply, etc.), the sentiment valence of their comment, as measured by VADER \cite{hutto2014vader}; the overall score of the comment (number of upvotes $-$ number of downvotes computed when Pushshift collected those posts); the timestamp of the comment relative to the subreddit's start and end date; four features used by Russo et al. \citeyear{russo2024stranger} representing the activity level of the thread prior to the newcomer's comment (number of comments, number of unique comments, number of top-level comments, and number of unique top-level commenters); and four features representing each user's activity on Reddit prior to posting in their respective hate subreddit (number of comments, number of submissions, number of unique threads posted in, and the net number of upvotes minus downvotes received on all comments and submissions). We also show in the appendix (Fig.~\ref{fig:cosine_similarity_matching}) that the text embeddings were similar across control and treatment, confirming that they say semantically similar things. We also present examples of matched newcomer triplets and their replies in the appendix (Table~\ref{tab:triplet_examples}). Importantly, none of the features used for matching include information from after the newcomer's comment was created.

In addition to matching users within hate subreddits, we matched users within non-hate subreddits. Because we did not measure counterspeech in non-hate subreddits, we used a binary treatment variable, optimally matching newcomers who received a reply to those who did not receive a reply. Since the percentage of newcomers who receive a reply is approximately equal to those who do not receive a reply (avg. percentage = 40\%), we use a random 10\% subsample of users who receive a reply as our treatment group, ensuring there are enough control users for each treatment user to be matched with a sufficiently similar control user. 

After collecting all the features used to match users, we experimented with two different ways of representing the features before using them for matching. The first method involved training a logistic regression propensity score model, where the input is a concatenated vector of all of the features (776 dimensions, with 768 of these coming from the text embedding). In this arrangement, each user was assigned a vector of length three, representing the probability from the propensity score model that the user belonged to each treatment group. As an alternative to propensity score matching, we performed principal component analysis on the input features, reducing them down to five dimensions (the average variance explained across all subreddits is $77\%$). We used Mahalanobis distance to calculate the distance between users. In preliminary analyses, we also considered training propensity score models using BERT (as done in \citet{russo2024stranger}), but we found this approach to be difficult to scale up due to the time needed to train BERT-based models. Furthermore, initial results suggested that the BERT-based approach is more sensitive to hyperparameters, and seemed to perform worse than a logistic regression model. We evaluated the quality of each matching approach by calculating the absolute standardized mean difference for each covariate for each pair of treatment groups. We show the absolute standardized mean differences in the Appendix Figure~\ref{fig:covariate_balance}, which demonstrates that the best approach uses a logistic regression propensity score model, since all covariates are either under the recommended 0.1 threshold for SMD, or very close to it \cite{austin2011introduction}.

\subsubsection{Subreddit Matching}

To compare the overall effects of replies in hate subreddits to the effects of replies in non-hate subreddits, we needed a large set of likely non-hate subreddits to match the hate subreddits. We used the list of the 10k largest subreddits obtained by Waller \& Anderson \citeyear{waller2021quantifying} to source non-hate subreddits. We used Mahalanobis distance matching to pair each hate subreddit with a non-hate subreddit from this sample, employing the same process described for the user matching but with different features. We removed NSFW subreddits from the matching pool, since we found them to frequently appear in preliminary matchings to the point that they dominated our non-hate sample, and we did not want this categorically distinct type of subreddit to dominate our results.

Comparing the engagement in banned (hate) subreddits and non-banned subreddits can bias the results of our analysis because posts by users who would otherwise be active after a subreddit is banned would not appear in the dataset. In this case, the engagement of users might appear lower simply because the posts they would have made could not be created. We accounted for this when matching subreddits. Namely, for each hate subreddit, we obtained the 90th percentile of the time taken for newcomers to post a second time. This value, alongside the number of users within the subreddit, and the activity rate of the subreddit defined as the average number of posts per user per month, was used for matching the subreddits. Limiting to the 90th percentile prevented our results from being skewed by users who could not post a second time because the subreddit had been banned. Moving forward, a ban was simulated for each non-hate subreddit by excluding posts made after their respective matched hate subreddit's ban date. Additionally, users who posted after the 90\% percentile return time before the ban were excluded, as they may have been unable to continue posting due to the (simulated) ban. 

The quality of our matching method is shown in Appendix Figure~\ref{fig:subreddit_matching}. The absolute standardized mean difference of all matching variables is below the recommended 0.1 threshold \cite{austin2011introduction}, indicating sufficient quality.

\subsection{Estimating the Effects of Counterspeech}

We aimed to measure the effect of counterspeech on user engagement in hate subreddits, defining engagement as a binary variable, where a user is said to ``continue to engage'' in a given hate subreddit if they make a post or comment \emph{outside} the original thread in which they posted. The choice to define users who continue to engage as users who post within threads other than their original thread, rather than including users who post within the same thread, is important -- it could be that certain types of replies--including counterspeech--could elicit an argument from the newcomer, causing them to continue posting within the same discussion. However, through repeated turns within the same discussion, a user could be discouraged from participating in other discussions. Therefore, a user participating in multiple discussions within a hate community is a stronger indicator of continued interaction with the broader community.

In addition to measuring the effects of counterspeech alone, we are also interested in examining whether the toxicity of counterspeech plays a role in its efficacy. To measure toxicity, we used the toxicity attribute from the Perspective API \cite{Jigsaw2017}. The definition of toxicity used by Perspective is ``a rude, disrespectful, or unreasonable comment that is likely to make you leave a discussion.'' This model has been extensively validated and used for Reddit research in the past (e.g, \cite{russo2023spillover}). When measuring the effects of toxicity and comparing the toxicity levels of different types of replies, we used the direct toxicity probability obtained from the Perspective API. 

To understand how replies impact user retention in hate subreddits, we created a mixed-effect logistic regression model to measure the likelihood that a user continues to post in another thread after their first post. We fed in the subreddit as random effects and employed the features used to match users, as well as what kind of reply they received, and the toxicity of the reply, as fixed effects. We fit separate models for hate and non-hate subreddits, with counterspeech and hate speech not being measured in non-hate subreddits. To ensure that our models were valid, we computed the variance inflation factor (VIF) for each coefficient. VIF values greater than 5 are indicative of substantial multicollinearity in the model, potentially impacting results \cite{senaviratna2019diagnosing}. In cases in which we observed that the VIF was greater than 5, we removed the highest VIF features until all VIF values were less than 5. We ended up removing three out of the four features representing thread-level activity prior to the users commenting in the threads (all four features are highly correlated), retaining the feature representing the number of unique commenters in a given thread. We also removed the user activity feature representing the number of unique threads in which each user posted before participating in a hate subreddit, since it was highly correlated with the total number of comments each user made.

In addition to measuring the effects counterspeech and toxicity have on continued engagement within hate subreddits, we also measured the consequences that toxic counterspeech has on subsequent toxicity in the same discussion thread. To accomplish this, we measured the toxicity levels of all replies to the replies of newcomer comments. We then measured the probability of users receiving a toxic reply for different toxicity levels of counterspeech, hate speech, and other speech, where a toxic reply is defined as a reply with a toxicity probability higher than 0.7. We also experimented with a toxicity threshold of 0.5 and obtained qualitatively similar results.

%For robustness, we also measure the effect of counterspeech on users' overall activity levels within hate subreddits. Activity rates are defined as the number of posts/comments per week made by each user following their initial comment. We measure this at intervals of four, eight, and twelve weeks following a newcomer's initial comment. Similarly to how we measure the effect of counterspeech on user retention, we fit a mixed-effects negative binomial regression model to measure its effects.

\section{Results}

\subsection{Model Performance}

\begin{figure}
    \centering
    \includegraphics[width=0.9\columnwidth]{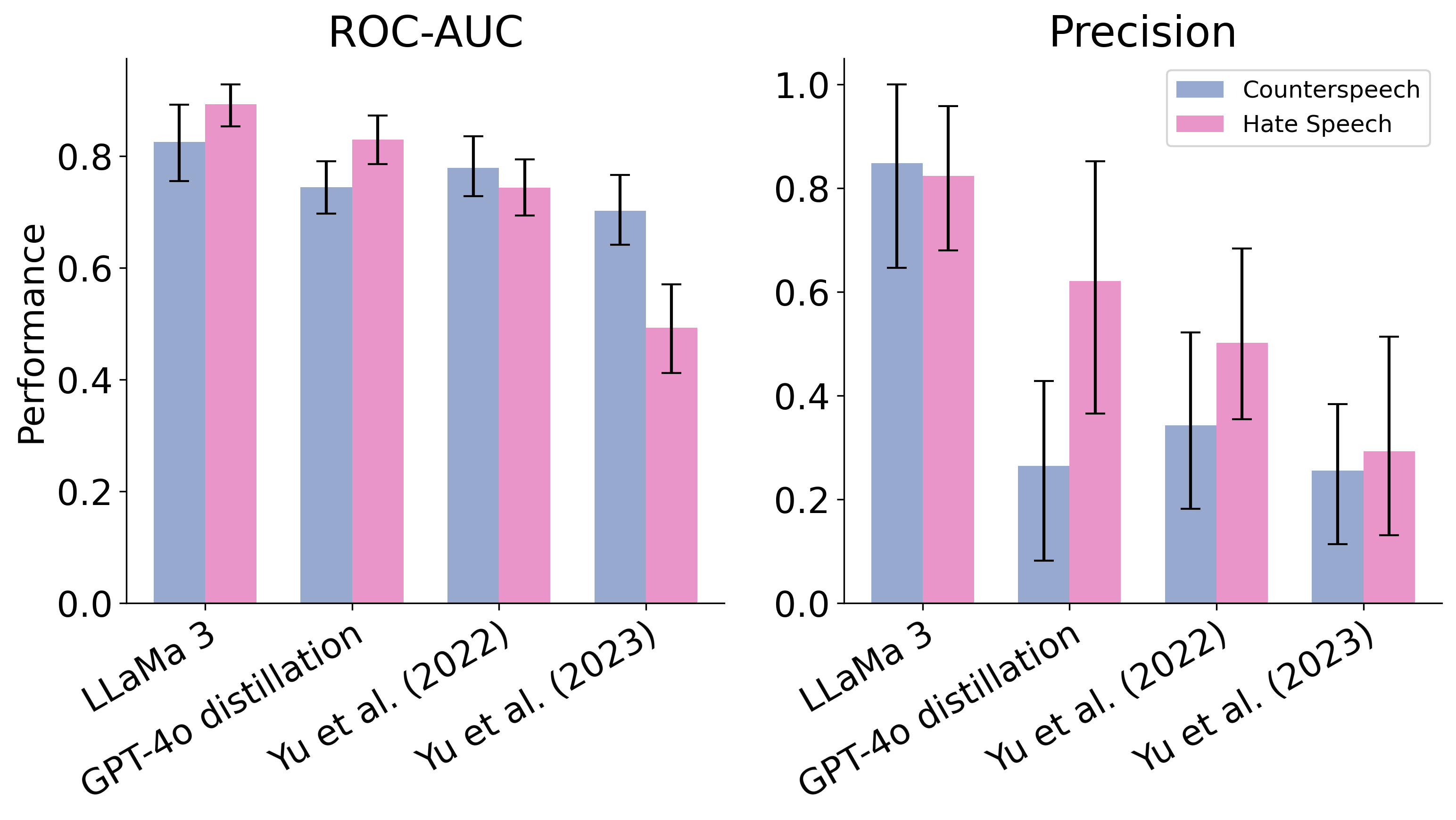}
    \caption{Performance of counterspeech detection models on the validation dataset. Black vertical lines represent 95\% confidence intervals obtained using bootstrapping on testing data. Precision values are obtained using the confidence threshold that maximizes precision while still retaining at least 10 predicted positives.}
    \label{fig:model_performance}
\end{figure}

Figure~\ref{fig:model_performance} shows the estimated precision and ROC-AUC values for a variety of different models. We noticed that while there are multiple models that achieved high ROC-AUC values, LLaMa 3.3 70B was the only model that achieved a suitable precision (mean precision is 0.85). When we fine-tuned RoBERTa on the dataset that we labeled using GPT-4o, or on the dataset produced by Yu et al. \citeyear{yu2022hate}, we found that the estimated precision values are considerably lower (0.26 and 0.34, respectively). Notably, for hate speech, our model trained on the dataset provided by Yu et al. \citeyear{yu2023fine} only achieved an estimated ROC-AUC value of 0.49, indicating that its predictions are no better than chance. Although the other RoBERTa models are better at predicting hate speech than counterspeech, they still do not achieve precision values as high as LLaMa 3.3 70B for hate speech detection. Given that both datasets obtained from prior literature were trained on Reddit, but in different communities from those that we studied, these results highlight the challenge of generalization for traditional supervised machine learning approaches, and add to the growing literature highlighting advantages that LLMs afford for generalizing across domains in computational social science tasks \cite{ziems2024can}.

\begin{table}[h]
\begin{tabular}{cccc}
\textbf{Confidence} & \textbf{\begin{tabular}[c]{@{}c@{}}CS\\ Precision\end{tabular}} & \textbf{\begin{tabular}[c]{@{}c@{}}HS\\ Precision\end{tabular}} & \textbf{\begin{tabular}[c]{@{}c@{}}Matched\\ Triplets\end{tabular}} \\ \hline
1.0                 & 0.88 $\pm$ 0.03                                                              & 0.92 $\pm$ 0.03                                                            & 12,473                                                                   \\
0.8                 & 0.81 $\pm$ 0.04                                                              & 0.91 $\pm$ 0.03                                                            & 16,513                                                                   \\
0.6                 & 0.74 $\pm$ 0.04                                                              & 0.90 $\pm$ 0.03                                                            & 21,652                                                                  
\end{tabular}
\caption{Final precision of counterspeech (CS) and hate speech (HS). Uncertainties represent standard errors.}
\label{tab:final_precision}
\end{table}

Although the performance of our LLaMa-based approach seemed initially promising given the results on the validation dataset, the confidence interval of the maximum precision is rather wide, and it is possible that we are overfitting to the validation dataset. To ensure that our model is sound, we annotated 100 predicted positive examples at three different confidence thresholds: 1, 0.8, and 0.6. The precision values of the thresholds, along with the sample sizes obtained by them, are reported in Table~\ref{tab:final_precision}. As expected, the precision of both hate speech and counterspeech is highest at the most restrictive threshold (especially for counterspeech), though the sample size is much smaller than at more relaxed thresholds. Here, we employ a threshold of 0.8, which we view as a reasonable compromise between accuracy and the ability to obtain a meaningful sample. Our results are largely robust no matter what threshold is used (see Appendix Figure~\ref{fig:regression_robustness_checks}).

\subsection{RQ1: Counterspeech is Highly Toxic}

\begin{figure}[h]
    \centering
    \includegraphics[width=0.8\columnwidth]{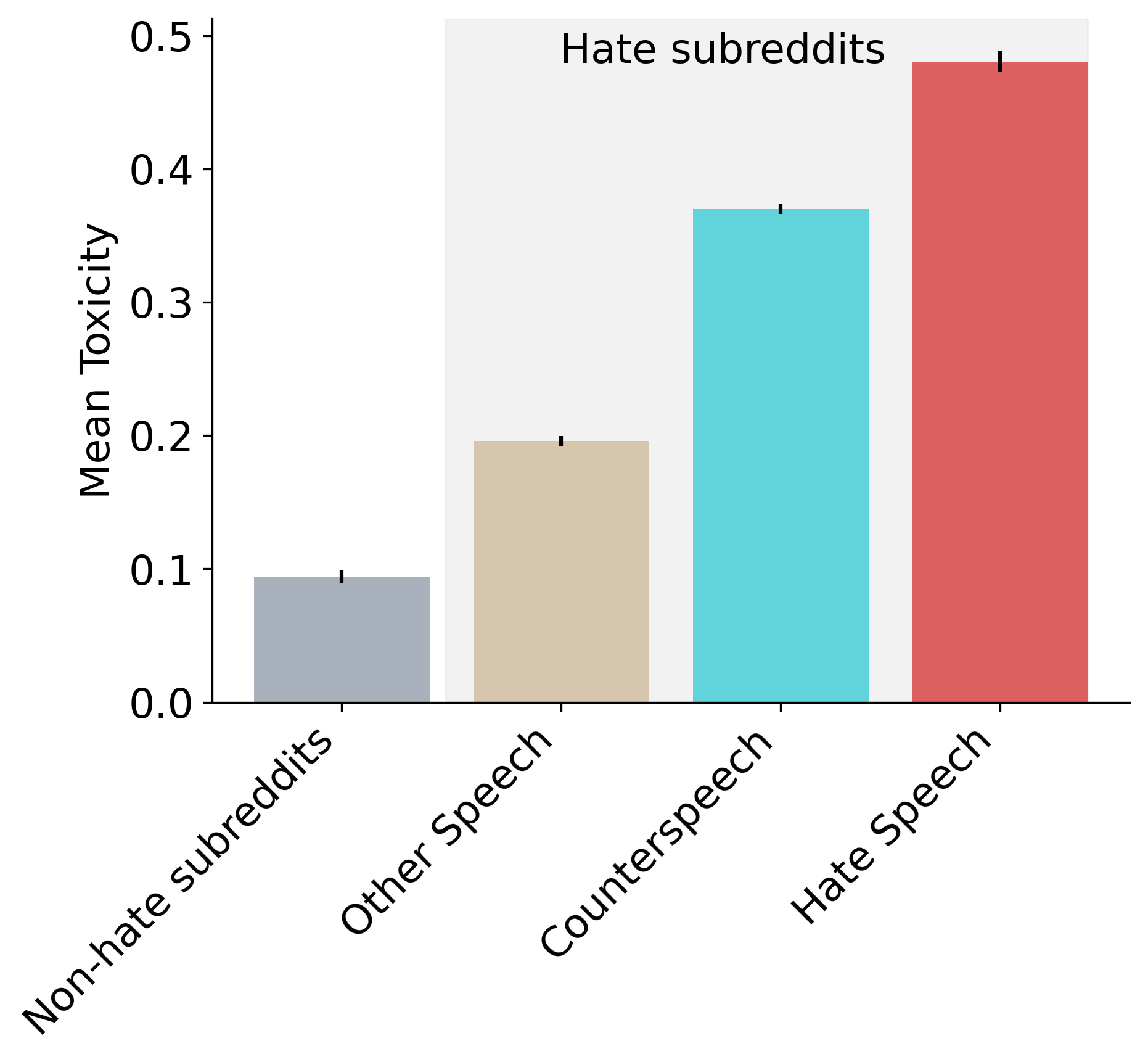}
    \caption{Mean toxicity by type of reply. Black vertical lines represent standard errors.}
    \label{fig:mean_toxicity}
\end{figure}

Figure~\ref{fig:mean_toxicity} shows the average toxicity probabilities of different types of replies. We find that, on average, replies detected as hate speech exhibited the highest toxicity levels, followed by replies detected as counterspeech. In hate subreddits, replies detected as counterspeech were approximately 88\% more toxic than replies detected as neither counterspeech nor hate speech, and 330\% more toxic than the average reply from non-hate subreddits.

Informed by prior research showing that counterspeech will often be misclassified as hate speech due to the mention of slurs in counterspeech \cite{gligoric2024nlp}, we considered the possibility that the high toxicity levels of counterspeech could be explained by counterspeakers quoting hateful users or mentioning slurs as part of their efforts to argue that such language is not acceptable. To ensure that our toxicity results are not due to this error, we repeated our analysis by removing quoted text from the replies containing counterspeech, and replacing common slurs with neutral terms. More details about this process can be found in the appendix. Appendix Figure~\ref{fig:toxicity_of_replies_adjusted} shows the adjusted mean toxicity values after performing this step, which are extremely similar to the unadjusted values.

\subsection{RQ2: Relationship Between Counterspeech and User Retention in Hate Communities}

\begin{figure}
    \centering
    \includegraphics[width=0.9\columnwidth]{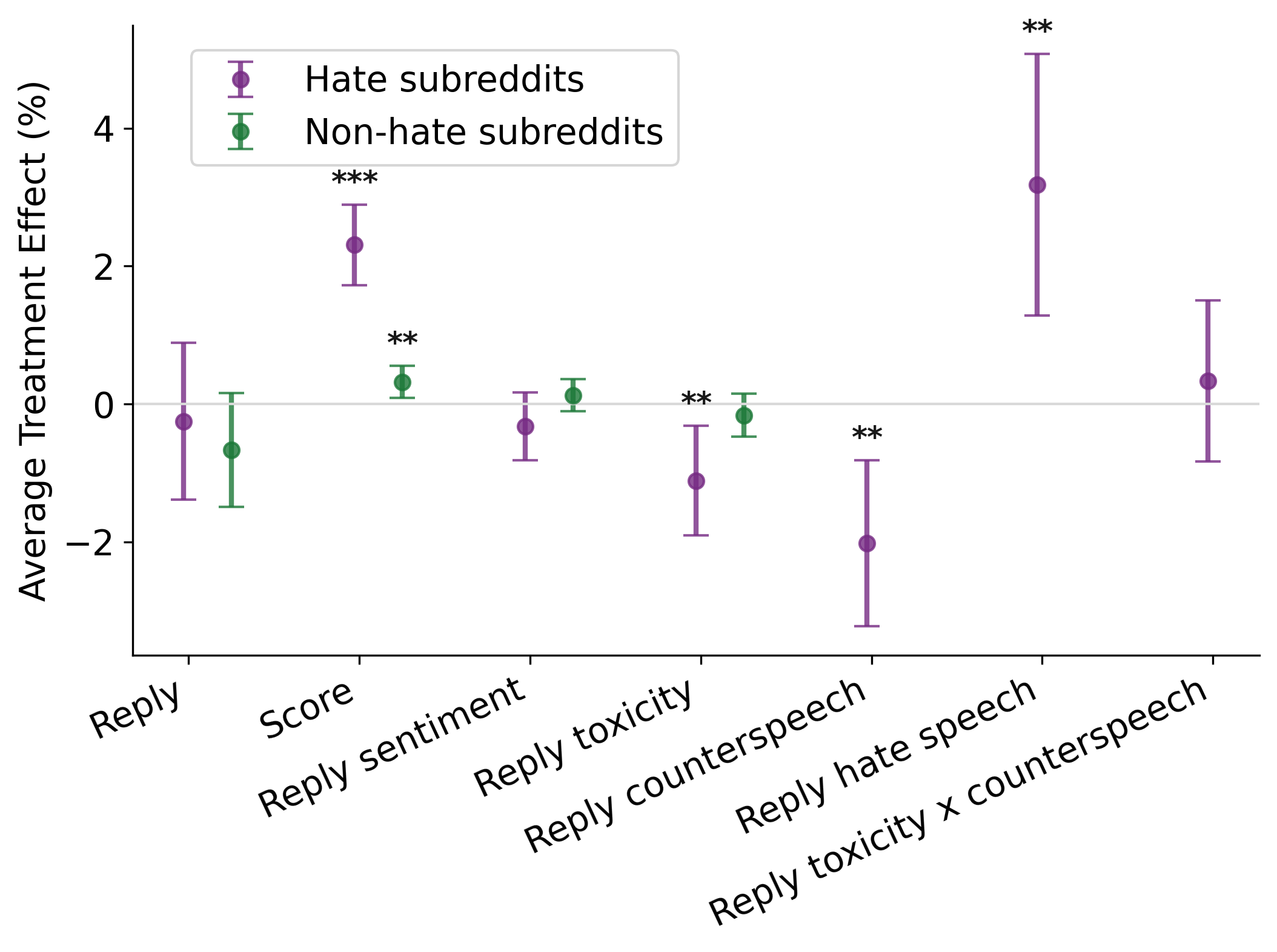}
    \caption{Average treatment effects of counterspeech, toxicity, and other key variables. Error bars represent 95\% confidence intervals.}
    \label{fig:treatment_effects}
\end{figure}

Figure~\ref{fig:treatment_effects} shows the output of our regression model. We found that counterspeech was negatively associated with newcomers' continued participation in hate groups (average treatment effect = -2.0\%, $p=0.001$). Meanwhile, replies containing hate speech were associated with increased engagement (average treatment effect = 3.1\%, $p = 0.001$). In both hate and non-hate subreddits, the total score of the newcomer's comment was associated with increased engagement, though the effect was higher for hate subreddits ($p < 0.001$), potentially indicating that social approval plays a larger role in users' adherence to hate groups \cite{walther2024effects}. Finally, toxicity did not have an effect on engagement for either category of subreddit. For hate subreddits, the interaction between toxicity and counterspeech also does not have a significant effect, i.e., we did not find evidence of a significant relationship between the toxicity of counterspeech and its effects.

To further investigate the relationship between toxicity and the effects of counterspeech, we repeated our analysis two additional times: once only including newcomers (and their matched counterparts) who received non-toxic counterspeech replies, and once only including newcomers (and their matched counterparts) who received toxic counterspeech replies. Following Perspective's official guidance, we used a toxicity threshold of 0.7 to differentiate between toxic and non-toxic replies\footnote{\url{https://developers.perspectiveapi.com/s/about-the-api-score?language=en_US}}. When performing the analysis on toxic replies, we found that the negative effect of counterspeech disappeared, and the result was no longer significant (ATE=12.7\%, $p=0.13$). However, counterspeech continued to have a negative effect on engagement when toxic counterspeech was removed from the sample (ATE=-2.0\%, $p<0.001$). The results were similar when using a toxicity threshold of 0.5 instead of 0.7. These results indicate that, while it is uncertain what the effects of toxic counterspeech are on engagement, non-toxic counterspeech does appear to be effective at reducing engagement in hate subreddits.

While the results presented in Figure~\ref{fig:treatment_effects} show significant effects of counterspeech, they were obtained using a specific classification threshold. Choosing a classification threshold to use for analysis is non-trivial, since there is a tradeoff between precision and recall: if our threshold is more restrictive, we have high precision, but a sample that is more biased to the types of counterspeech we are more confident in. With a less restrictive threshold, we will have a larger sample size, but with more false positive. Similarly, the type of matching approach we use could have an impact on the final results, despite our matching methods being roughly equal in performance. To ensure that our results are robust, we repeated our analyses using different counterspeech/hate speech thresholds, different matching algorithms, and measuring activity rates as the outcome variable. As shown in the Appendix (Fig~\ref{fig:regression_robustness_checks}, Fig~\ref{fig:robustness_sample_size}), our results on the effects of counterspeech were qualitatively similar under all conditions, although the results on the effects of toxicity are mixed.

\subsection{RQ2: Effects of Toxic Counterspeech Within the Same Discussion}

\begin{figure}
    \centering
    \includegraphics[width=0.8\columnwidth]{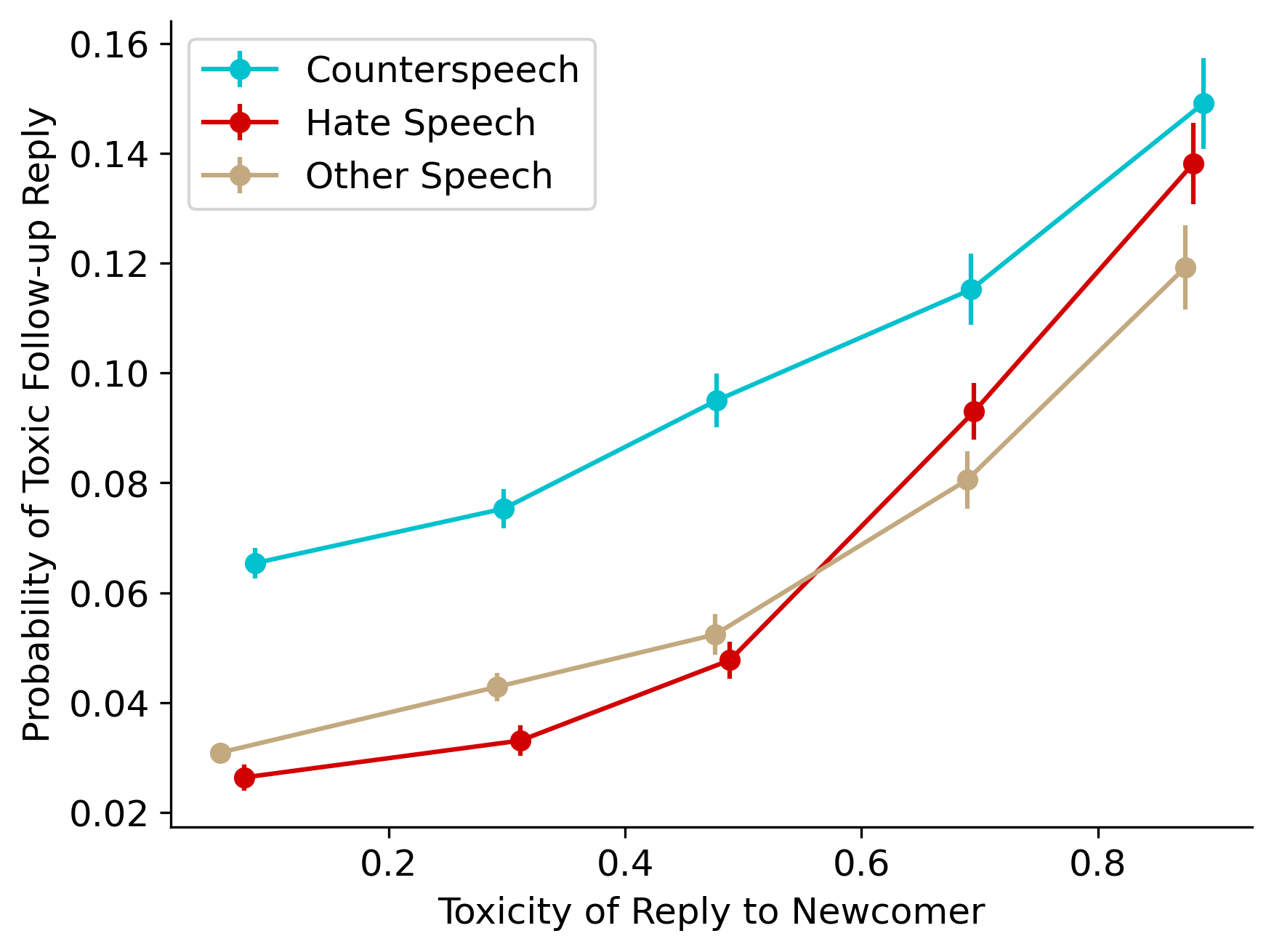}
    \caption{Probability of receiving a toxic follow-up reply based on toxicity levels of replies to newcomers. Points represent binned toxicity values. Vertical lines represent standard errors.}
    \label{fig:toxic_follow_ups}
\end{figure}

Figure~\ref{fig:toxic_follow_ups} shows the probability of different types of replies to newcomers eliciting a toxic follow-up reply. From the figure, it is apparent that, for all types of speech, the more toxic the reply to the newcomer, the more likely that reply is to receive a toxic follow-up. Additionally, counterspeech at lower toxicity levels is more likely to elicit toxic replies than hate speech or other speech at low toxicity levels, though highly toxic hate speech and highly toxic counterspeech exhibit similar rates of toxic follow-up replies. These results indicate the potential for toxic counterspeech to increase hostility within online discussions.

\section{Discussion}

To assess the effect that counterspeech has on engagement in hate forums, we analyzed replies to newcomer comments in a matched sample of 104 hate and 104 non-hate subreddits. We found that counterspeech was more likely to cause newcomers to leave hate subreddits than to continue engaging with those communities. We also found that counterspeech was generally toxic. Although we did not find a significant relationship between the toxicity of counterspeech and its effectiveness at deterring long-term engagement, toxic counterspeech increased the toxicity of responses within the same discussion thread. Furthermore, we cannot say whether, when users disengage from hate communities on Reddit, this is because they have changed their beliefs, or merely because they have changed their online behavior (potentially including moving to more hate-friendly platforms, etc.). Broadly, these results highlight the importance of further efforts to examine the effects of counterspeech, and to carefully weigh the pros and cons of different types of responses to hate speech. Simultaneously, these findings suggest that newcomers to hate groups might be more responsive to opposing views than might be presumed based on the antagonist nature of their hate speech. Below we discuss the implications of these findings for online policymakers, community moderators, and researchers.

\subsection{RQ1: Toxicity of Counterspeech}

While we find no statistically significant relationship between the toxicity of counterspeech and its effectiveness in deterring the targeted user from engaging in future discussions in a hate subreddit (Fig.~\ref{fig:treatment_effects}), the prevalence of toxic counterspeech is still cause for concern (Fig~\ref{fig:mean_toxicity}), as such toxicity can be expected to erode online environments in general. While counterspeech dissuades newcomers from engaging in hate groups beyond the context of their initial comment thread, newcomers are nevertheless more likely to reply to toxic counterspeech with incivility at first (Fig.~\ref{fig:toxic_follow_ups}), corroborating prior research demonstrating how toxicity begets more toxicity in online discussions \cite{yu2024hate, mann2003evolution, saveski2021structure, kim2021distorting}. Furthermore, repeated exposure to hostile exchanges may have spillover effects beyond the immediate discussion, shaping participants’ behavior and expectations in subsequent interactions and contributing to broader norm erosion across online communities.

Just as the evidence on the diffusion of toxic language on social media raises concerns about toxic counterspeech, it also offers a potential explanation for why we see such a high prevalence of toxicity in counterspeech, as hate speech is highly toxic (Fig.~\ref{fig:mean_toxicity}), and counterspeakers may be inclined to meet hate speech with toxicity. Furthermore, some evidence indicates that greater exposure to toxicity increases engagement within social media platforms, and hence platforms may have a financial incentive to uprank toxic content \cite{beknazar2025toxic}. Therefore, it could be that the design of online platforms plays a role in facilitating highly toxic counterspeech. For this reason, better platform design and the development of interactive tools intended to help counterspeakers compose effective responses hold great promise as potential solutions to improve the quality of online counterspeech \cite{ding2024counterquill, ping2025perceiving}. For example, rather than treating counterspeech as a purely organic intervention, platforms could support counterspeakers through design affordances that encourage reflection and reduce escalation, such as pre-reply prompts, reframing suggestions, or reduced amplification of hostile responses. In this sense, moderation and counterspeech need not be competing strategies: moderation can address clear violations of platform rules, while well-designed counterspeech tools can help users contest hate in ways that limit toxicity and reduce downstream harm.

\subsection{RQ2: Counterspeech Of All Forms Reduces Engagement}

While previous results highlight the promise of counterspeech in reducing hate speech on X \cite{he2021racism, hangartner2021empathy}, one might expect that, within the context of a hate group, counterspeech is less likely to dissuade users from participating, as they are in an environment where counterspeech represents the minority opinion -- and prior investigations of cross-partisan interactions show that such exchanges fail to have a depolarizing effect \cite{xia2025integrated, bail2018exposure}. Surprisingly, we find that the opposite is true: newcomers to hate subreddit who receive replies containing counterspeech are less likely to continue engaging in the given hate subreddit. Taken together, our results largely corroborate prior literature showing that counterspeech is effective at reducing hate, extending it to a different context. One possible explanation is that newcomers to hate subreddits are often at an early stage of engagement, before hate ideologies have become deeply embedded in their identities or social relationships within the community. While we acknowledge that our study cannot measure belief change directly, this conjecture is consistent with prior research suggesting that the adoption of counter-normative beliefs is typically a gradual process rather than an instantaneous one, often beginning with tentative or playful engagement \cite{luhrmann1991persuasions}. Correspondingly, scholars of online hate and extremism have argued that humor serves as a key recruitment mechanism, allowing users to interact with hateful ideas in a low-commitment or ambiguous manner before fully endorsing them \cite{schmid2025humor, Schmid12032025}. In this light, counterspeech may be particularly effective for newcomers because it intervenes at a point where participation and commitment remain reversible. To more thoroughly investigate this possibility, future work should explore more nuanced indicators of commitment to hate ideologies from text (e.g., Lahnala et al. \citeyear{lahnala2025unifying}), examining the relationship between these indicators and resistance to counterspeech. Indeed, research such as this could help to identify neophyte hate users who are likely to be persuaded by counterspeech.

While reduced engagement does not necessarily imply belief change, disengagement may nevertheless have important downstream consequences that extend beyond the immediate interaction. One possibility is a shift toward passive consumption rather than continued participation, which can still reduce the public visibility, social reinforcement, and diffusion of hateful content, even if private beliefs remain unchanged. Because posting and interaction play a central role in consolidating group identity and escalating commitment, reduced participation may interrupt social feedback loops that sustain hate communities \cite{cheng2015formation}. At the same time, exposure to counterspeech may increase the perceived social or normative cost of participation, prompting some users to migrate to other platforms or communities perceived as more ideologically aligned or less contested -- a dynamic that parallels displacement effects observed following moderation interventions \cite{horta2021platform}. Distinguishing among these downstream trajectories remains an important direction for future research.

\subsection{Limitations and Future Directions}
While we investigated the effects of counterspeech on engagement in hate subreddits to the best of our ability, there remain additional challenges and questions about recruitment in online hate groups. First, while the models we use have state-of-the-art performance for the dataset studied, other models, if made equally accurate, might produce different results. Relatedly, while we did explore whether a failure to distinguish between \textit{use} of hate speech and \textit{mention} of hate speech (as when a counterspeaker quotes a hate group member) could have caused the Perspective API to misclassify counterspeech as toxic, the Perspective API could also misclassify posts in other biased ways for which we did not, such as inaccurately detecting certain dialects as toxic \cite{sap2022annotators}. In addition to limitations of prediction models, there are relevant outcomes beyond user retention that we did not (or could not) examine, such as whether users migrate to other platforms after receiving counterspeech replies, or continue to spread hate speech outside of hate subreddits. Furthermore, future research could expand the population of users studied to established members of hate communities to more directly compare them to newcomers. Additionally, given the limited number of platforms on which studies of counterspeech have been conducted, it is important that research on counterspeech consider a broader view of social media, including alt-tech platforms, as well as platforms, such as TikTok, that have more recently become more popular. Future research can also investigate community-level attributes that are predictive of the efficacy of counterspeech, and assess the utility of a more fine-grained taxonomy of counterspeech.

\section{Ethical Statement}
A number of ethical considerations apply to this study. First, while we analyzed the impact of (surprisingly common) toxic counterspeech, it is important to underscore that we are not advocating toxic speech of any form. Indeed, toxicity likely adds to the broader negativity in online discourse, and our results point to its lack of additional efficacy compared to non-toxic counterspeech. Next, we recognize that there are ways this work could also be misused, such as by members of hate groups to persuade outsiders, but we believe this risk is minimal. In addition, there can be a significant harm if these models are used to single out or punish individual users, especially given the possibility that the models may misclassify speech. We therefore emphasize that the models should only be used to understand aggregate patterns. Finally, our work leaves open potential ethical questions, such as who should be responsible for creating counterspeech, and whether it can add additional burdens on moderators.

\section{Acknowledgements}

The authors would like to thank Hillel Cogan, Derrick Liu, Daniel Penn, Rafael Arellano, and Dan Faltesek for their valuable work annotating counterspeech and hate speech. ChatGPT was used to generate code and tables for this project, as well as minor rephrasing of some sentences in the manuscript. All generative AI ouptuts have been verified by the authors.

%\bibliography{citations}

\subsection{Paper Checklist}
\begin{enumerate}

\item For most authors...
\begin{enumerate}

    \item  Would answering this research question advance science without violating social contracts, such as violating privacy norms, perpetuating unfair profiling, exacerbating the socio-economic divide, or implying disrespect to societies or cultures?
    \answerYes{Yes.}
  \item Do your main claims in the abstract and introduction accurately reflect the paper's contributions and scope?
   \answerYes{Yes.}
   \item Do you clarify how the proposed methodological approach is appropriate for the claims made? 
   \answerYes{Yes.}
   \item Do you clarify what are possible artifacts in the data used, given population-specific distributions?
    \answerYes{Yes, see Methods.}
  \item Did you describe the limitations of your work?
    \answerYes{Yes, see Limitations.}
  \item Did you discuss any potential negative societal impacts of your work?
    \answerYes{Yes, see Ethical Considerations.}
  \item Did you discuss any potential misuse of your work?
    \answerYes{Yes, see Ethical Considerations.}
    \item Did you describe steps taken to prevent or mitigate potential negative outcomes of the research, such as data and model documentation, data anonymization, responsible release, access control, and the reproducibility of findings?
    \answerYes{Yes, see Methods, as well as documentation/code in the code repository.}
  \item Have you read the ethics review guidelines and ensured that your paper conforms to them?
   \answerYes{Yes.}
\end{enumerate}
\item Additionally, if your study involves hypotheses testing...

\begin{enumerate}
  \item Did you clearly state the assumptions underlying all theoretical results?
     \answerNA{NA}
  \item Have you provided justifications for all theoretical results?
   \answerNA{NA}
  \item Did you discuss competing hypotheses or theories that might challenge or complement your theoretical results?
  \answerNA{NA}
  \item Have you considered alternative mechanisms or explanations that might account for the same outcomes observed in your study?
   \answerNA{NA}
  \item Did you address potential biases or limitations in your theoretical framework?
   \answerNA{NA}
  \item Have you related your theoretical results to the existing literature in social science?
  \answerNA{NA}
  \item Did you discuss the implications of your theoretical results for policy, practice, or further research in the social science domain?
 \answerNA{NA}
\end{enumerate}
\item Additionally, if you are including theoretical proofs...

\begin{enumerate}
  \item Did you state the full set of assumptions of all theoretical results?
 \answerNA{NA}
	\item Did you include complete proofs of all theoretical results?
 \answerNA{NA}
\end{enumerate}
\item Additionally, if you ran machine learning experiments...

\begin{enumerate}
  \item Did you include the code, data, and instructions needed to reproduce the main experimental results (either in the supplemental material or as a URL)?
    \answerYes{Yes, see the Introduction for data and code.}
  \item Did you specify all the training details (e.g., data splits, hyperparameters, how they were chosen)?
    \answerYes{Yes, and further details are within the attached code.}
     \item Did you report error bars (e.g., with respect to the random seed after running experiments multiple times)?
    \answerYes{Yes}
	\item Did you include the total amount of compute and the type of resources used (e.g., type of GPUs, internal cluster, or cloud provider)?
    \answerYes{Yes, see subsection Counterspeech Detection in the Methods section.}
     \item Do you justify how the proposed evaluation is sufficient and appropriate to the claims made? 
    \answerYes{Yes, see Methods.}
     \item Do you discuss what is ``the cost`` of misclassification and fault (in)tolerance?
    \answerYes{Yes, see Ethical Considerations.}
\end{enumerate}
\item Additionally, if you are using existing assets (e.g., code, data, models) or curating/releasing new assets, \textbf{without compromising anonymity}...
\begin{enumerate}
  \item If your work uses existing assets, did you cite the creators?
   \answerYes{Yes}
  \item Did you mention the license of the assets?
    \answerYes{Yes, the code is MIT licensed, as described in the repository link.}
  \item Did you include any new assets in the supplemental material or as a URL?
    \answerYes{Yes, we share code and data in our anonymized URL.}
  \item Did you discuss whether and how consent was obtained from people whose data you're using/curating?
   \answerYes{
   Yes, data was gathered without subject's consent as data are publicly avaialble and anonymized to remove any PII.}
  \item Did you discuss whether the data you are using/curating contains personally identifiable information or offensive content?
    \answerYes{Yes, no data contains PII.}
\item If you are curating or releasing new datasets, did you discuss how you intend to make your datasets FAIR (see \citet{fair})?
\answerYes{Yes, see Methods.}
\item If you are curating or releasing new datasets, did you create a Datasheet for the Dataset (see \citet{gebru2021datasheets})? 
\answerYes{Yes, see the code repository link.}
\end{enumerate}
\item Additionally, if you used crowdsourcing or conducted research with human subjects, \textbf{without compromising anonymity}...
\begin{enumerate}
  \item Did you include the full text of instructions given to participants and screenshots?
     \answerNA{NA}
  \item Did you describe any potential participant risks, with mentions of Institutional Review Board (IRB) approvals?
    \answerNA{NA}
  \item Did you include the estimated hourly wage paid to participants and the total amount spent on participant compensation?
    \answerNA{NA}
   \item Did you discuss how data is stored, shared, and deidentified?
  \answerNA{NA}
\end{enumerate}
\end{enumerate}

\section{Appendix}

\appendix
\renewcommand{\theequation}{S\arabic{equation}}
\renewcommand{\thetable}{S\arabic{table}}
\renewcommand{\thefigure}{S\arabic{figure}}
\setcounter{equation}{0}
\setcounter{table}{0}
\setcounter{figure}{0}

\begin{figure*}[ht]
    \centering
    \includegraphics[width=0.8\textwidth]{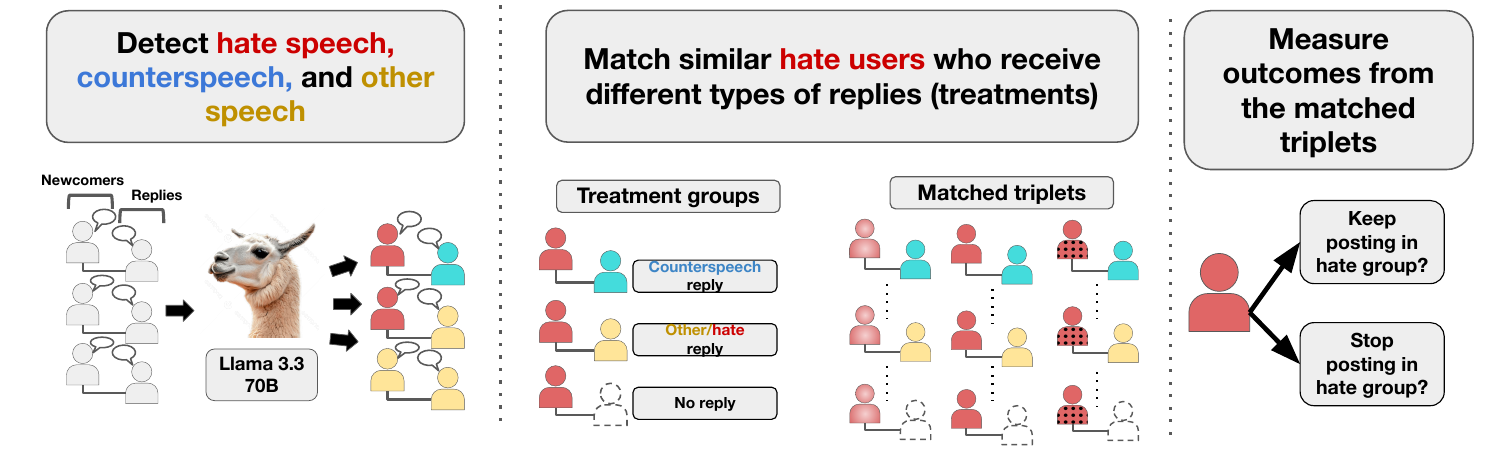}
    \caption{The impact of hate speech, counterspeech, and other speech on user retention in hate groups. We first collect all replies to new hate group users, and classify it as ``hate speech'', ``counterspeech'', or ``other'' using a fine-tuned Llama 3.3 70B model. We then use distance matching to compare similar hate group users in three treatment groups: those that receive one of the three replies after their first post, and compare against a control group: new hate group users that do not receive a reply after their first post. Finally we measure whether new user post activity changes any of the treatment conditions.}
    \label{fig:methods_schematic}
\end{figure*}

\begin{figure}
    \centering
    \includegraphics[width=0.9\columnwidth]{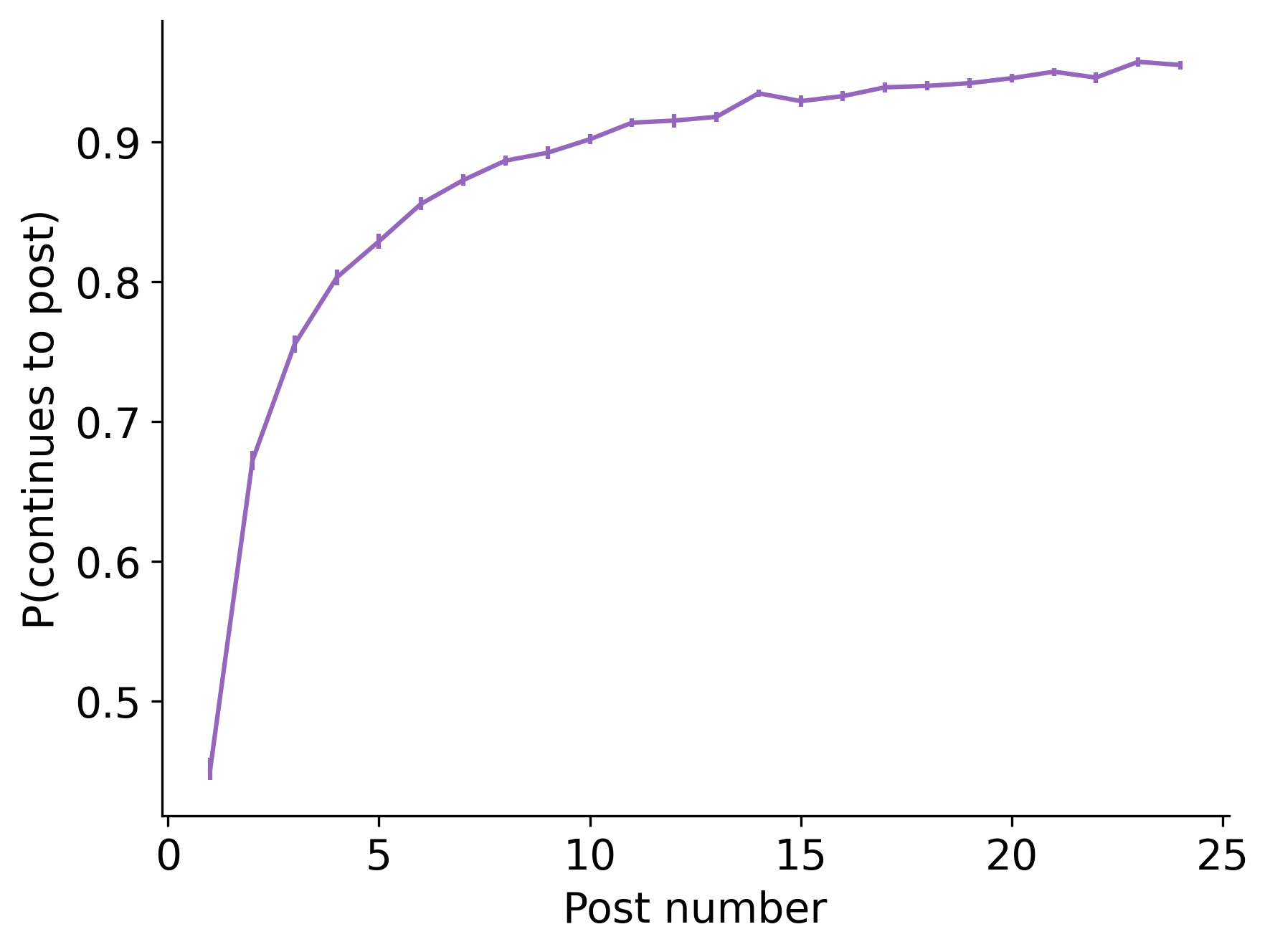}
    \caption{Average probability of continuing to post in a hate subreddit after posting $n$ times ($n$ is represented by the value on the x-axis). For example, the y-axis value for post number 5 represents the probability that a user who has posted in 5 different threads will post a 6th time. Vertical lines represent standard errors.}
    \label{fig:prob_continue}
\end{figure}

\begin{figure*}[h]
\centering % Center the textbox within the figure environment
\begin{tcolorbox}[colframe=purple!50, colback=purple!10, title={\textbf{Counterspeech prompt}},
]
    \small

You are a model that identifies whether a pair of comments on Reddit are using counterspeech, hate speech, or other speech. You provide your answers in the format (newcomer: prediction, reply: prediction). \\

Hate speech is defined by the United Nations as ``any kind of communication in speech, writing or behaviour, that attacks or uses pejorative or discriminatory language with reference to a person or a group on the basis of who they are, in other words, based on their religion, ethnicity, nationality, race, colour, descent, gender or other identity factor.'' Counterspeech is speech that opposes hate speech. `Other' speech is neither counterspeech nor hate speech.
\\

The comments come from a hate subreddit, so speech that calls out the community for being hateful or ideologies of hate may qualify as counterspeech. Note that counterspeech may still be hostile or contain insults, but as long as it is speaking out against harmful speech directed at identity groups, such as racism, misogyny, anti-LGBTQ speech, or ableism, it will still qualify as counterspeech. \\

You can assign one of three labels to each comment. Note that the comments may each have a different label:\\
``Other''\\
``Hate speech''\\
``Counterspeech''\\

You will be given three comments: ``context,'' ``newcomer,'' and ``reply.'' The ``newcomer'' comment is a reply to the ``context'' comment, and the ``reply'' is a reply to the “newcomer” comment. Your goal is to identify whether the ``newcomer'' and ``reply'' comments are neutral, hate speech, or counterspeech. Please provide your answer in the format of (newcomer: prediction, reply: prediction). Do not include any additional details in your reply.\\

Context: (HATE SPEECH)\\
Newcomer: (HATE SPEECH)\\
Reply: (OTHER SPEECH)\\
Answer: (Newcomer: hate speech, reply: other)\\

(four more examples)

Context: [CONTEXT COMMENT HERE]\\
Newcomer: [NEWCOMER COMMENT HERE]\\
Reply: [REPLY HERE]\\
Answer:

\end{tcolorbox}
    \caption{Prompt for detecting counterspeech, hate speech, and other speech. We omit the few-shot examples we used, since they contain hate speech. The prompt including few-shot examples can be found in our GitHub repository.}
    \label{fig:prompt}
\end{figure*}

\begin{figure}
    \centering
    \includegraphics[width=0.9\columnwidth]{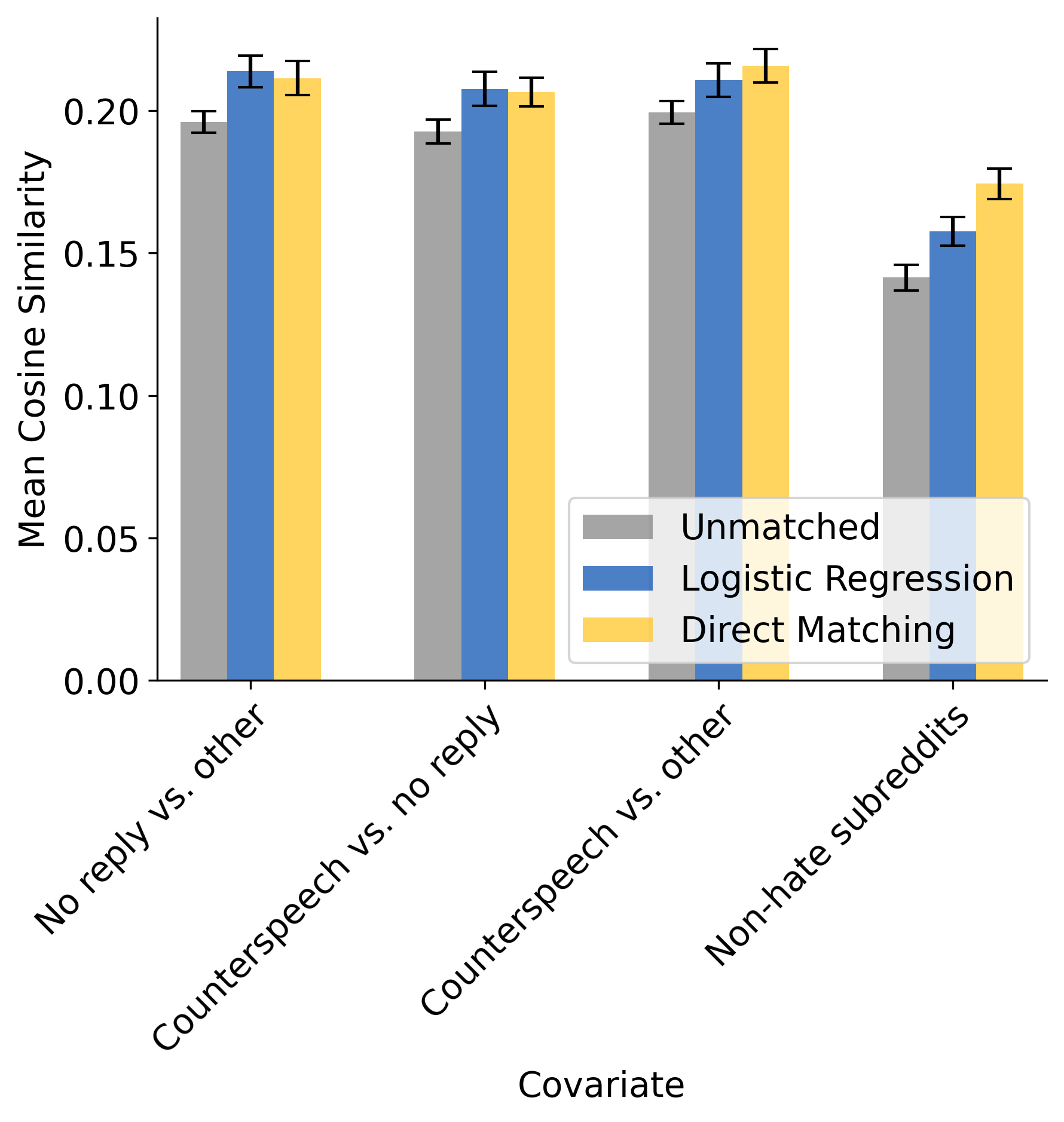}
    \caption{Average cosine similarity of matched users for different matching approaches.}
    \label{fig:cosine_similarity_matching}
\end{figure}

\begin{figure*}
    \centering
    \includegraphics[width=0.9\textwidth]{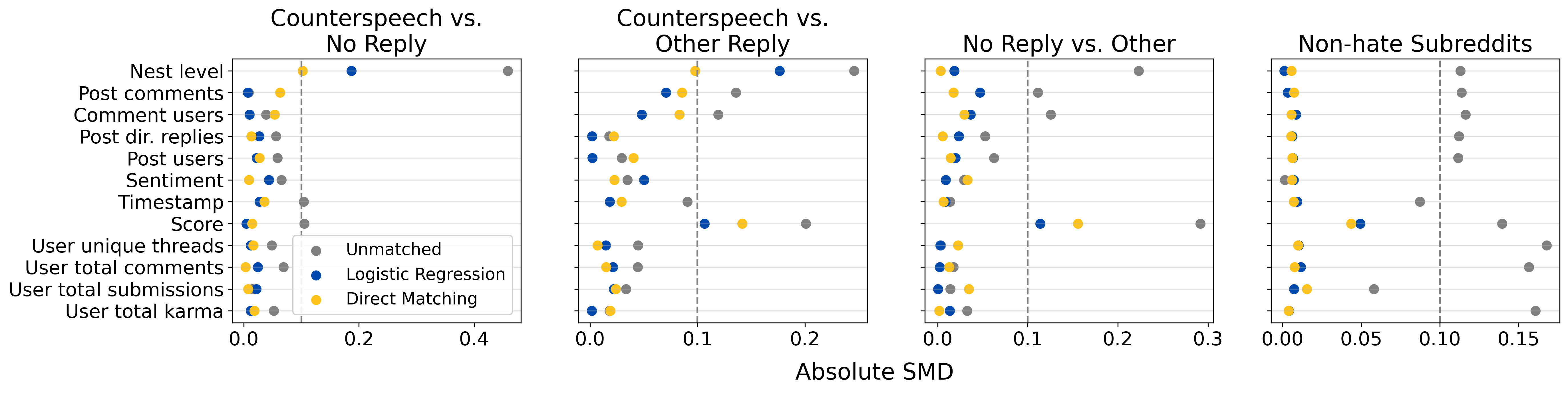}
    \caption{Absolute standardized mean differences of features used to match newcomers of different treatment groups. Each subplot represents a different combination of treatments. The dashed gray line represents an absolute standardized mean difference of 0.1 (the recommended threshold).}
    \label{fig:covariate_balance}
\end{figure*}

\begin{figure}
    \centering
    \includegraphics[width=0.8\columnwidth]{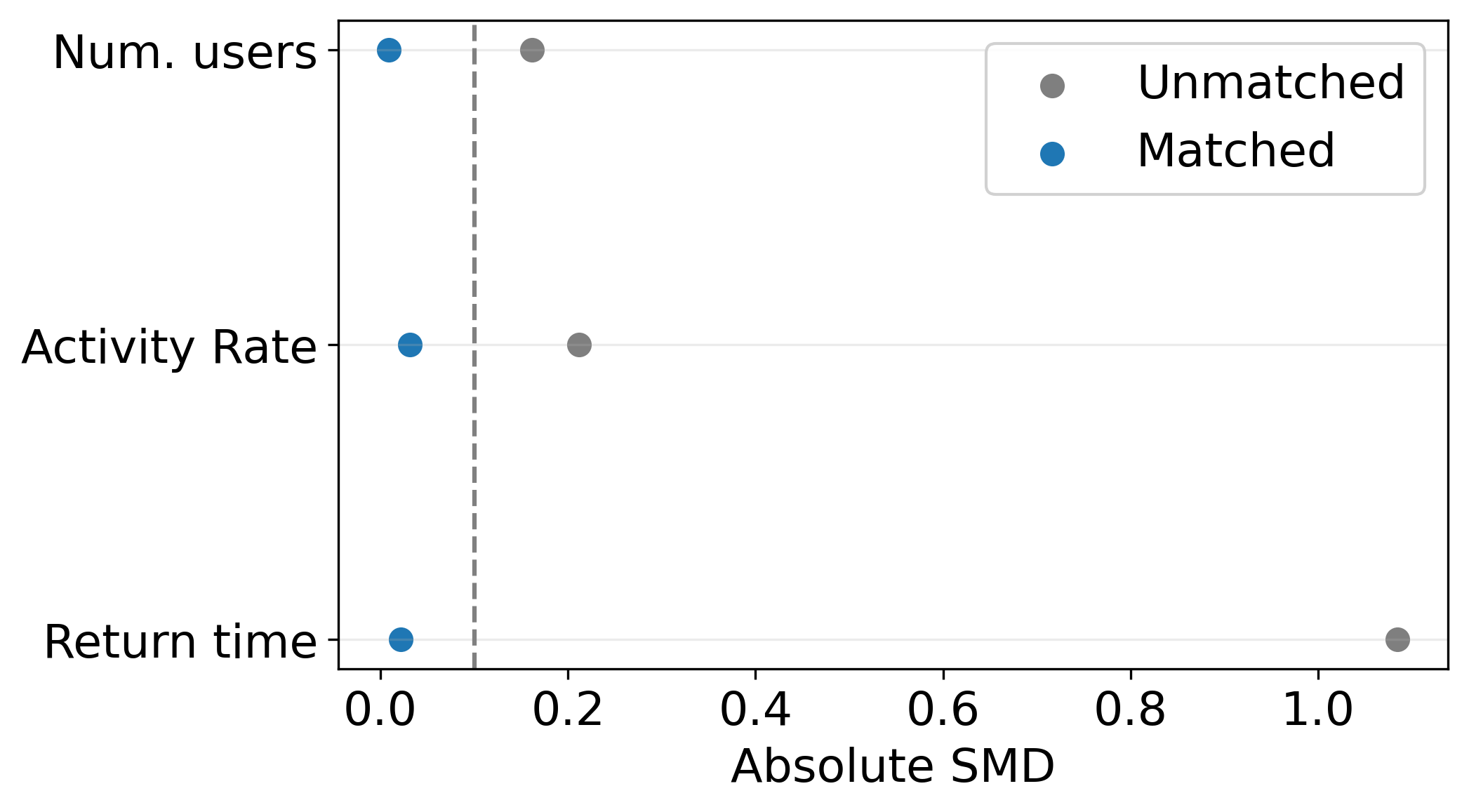}
    \caption{Absolute standardized mean differences of features used to match hate subreddits with non-hate subreddits. The dashed gray line represents an absolute standardized mean difference of 0.1.}
    \label{fig:subreddit_matching}
\end{figure}

\subsection{Toxicity of Counterspeech}

Prior research on hate speech detection shows that counterspeech is often misclassified as hate speech, because hate speech detection models fail to capture the ``use-mention'' distinction: sometimes, a counterspeaker may include a slur in their comment, but only ``mention'' it instead of use it. For example, they might mention the slur to tell someone that it is not acceptable to use that word, or they might quote specific parts of a hate comment in order to more directly respond to each point \cite{gligoric2024nlp}. Given these possibilities, it is plausible that the high toxicity we observed in counterspeech is explained by this distinction, since counterspeech that mentions hate speech would be detected as more toxic. To understand whether the use-mention distinction plays a role in skewing our results, we first removed all block quotes from the counterspeech replies (indicated by the HTML entity ``\&gt;''), then we replaced common slurs with the word ``person.'' To identify slurs that would frequently appear in counterspeech, we used TF-IDF to find the top 1000 keywords most characteristic of the counterspeech replies, after removing stopwords. In the list of keywords, we only found a few misogynistic slurs. Therefore, we constructed a list of slurs we expected to appear more frequently than other slurs, including the racist n-word, the homophobic f-slur, the transphobic t-slur, the antisemitic k-slur, and a few other misogynistic slurs. The full list of slurs can be found in our GitHub repository. We replaced each of these terms with the word ``person'' and recalculated the toxicity levels of counterspeech replies with quotes removed and slurs replaced.

\begin{figure}
    \centering
    \includegraphics[width=0.9\columnwidth]{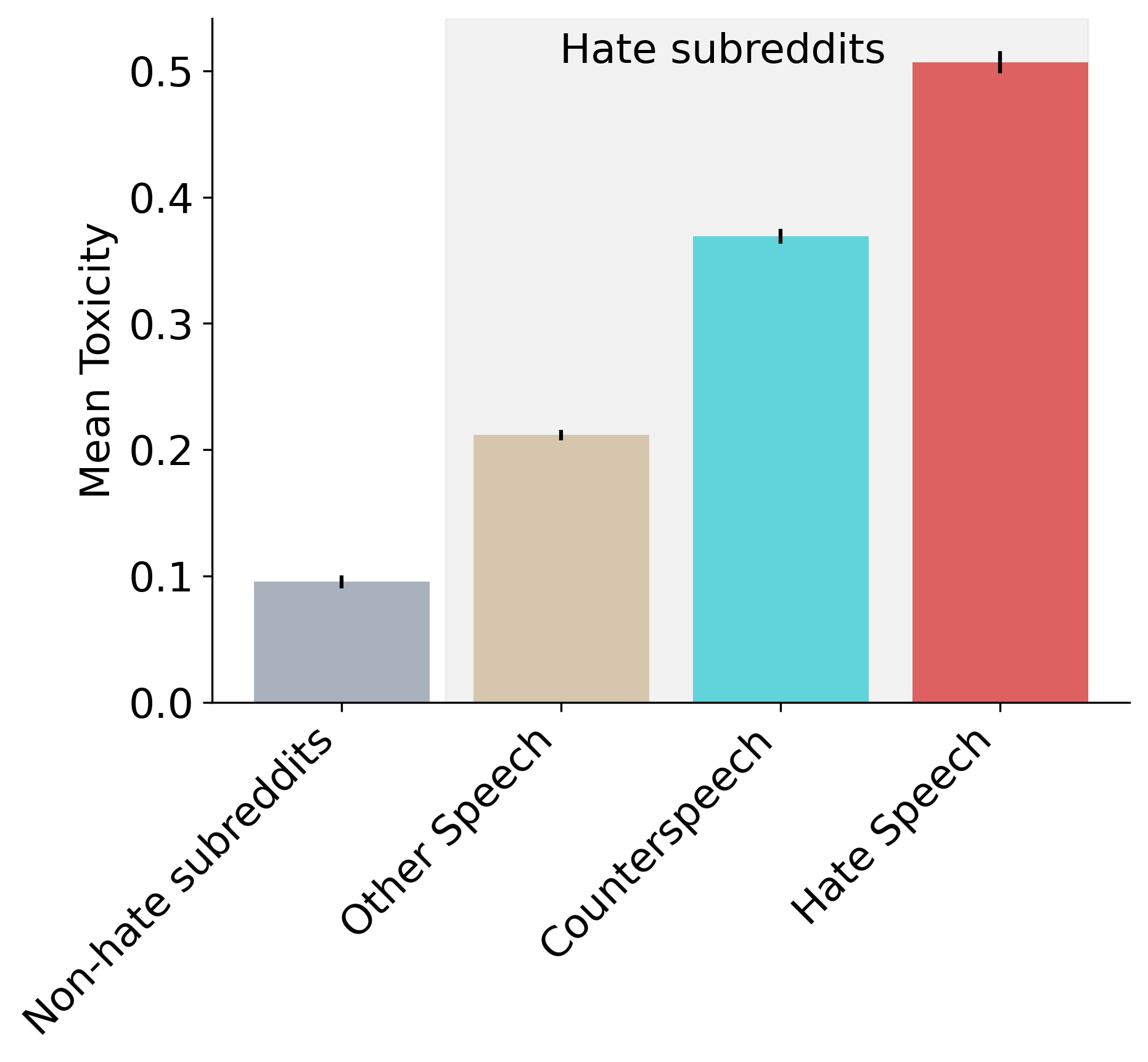}
    \caption{Mean toxicity by reply type after adjusting for common slurs. The counterspeech estimate is adjusted such that common slurs are replaced with neutral terms and block quotes are removed from the text.}
    \label{fig:toxicity_of_replies_adjusted}
\end{figure}

Figure~\ref{fig:toxicity_of_replies_adjusted}
shows the mean toxicity levels of different types of replies, with the counterspeech replies adjusted to account for the use-mention distinction. From the figure, it is apparent that removing quoted text and replacing slurs did not qualitatively change the results, indicating that the mention of hate speech is likely too rare to dramatically skew the toxicity of our estimates.

\subsection{Measuring the Effects of Counterspeech: Robustness Checks}

To estimate the effects of counterspeech, we must make several decisions in the process to which there are not obvious answers. For example, choosing the confidence threshold to use for hate speech/counterspeech classification requires making an important tradeoff between precision and recall. In this section, we describe additional robustness checks that we performed to ensure the results of our study were consistent under equally valid modeling choices.

\begin{figure*}
    \centering
    \includegraphics[width=0.9\textwidth]{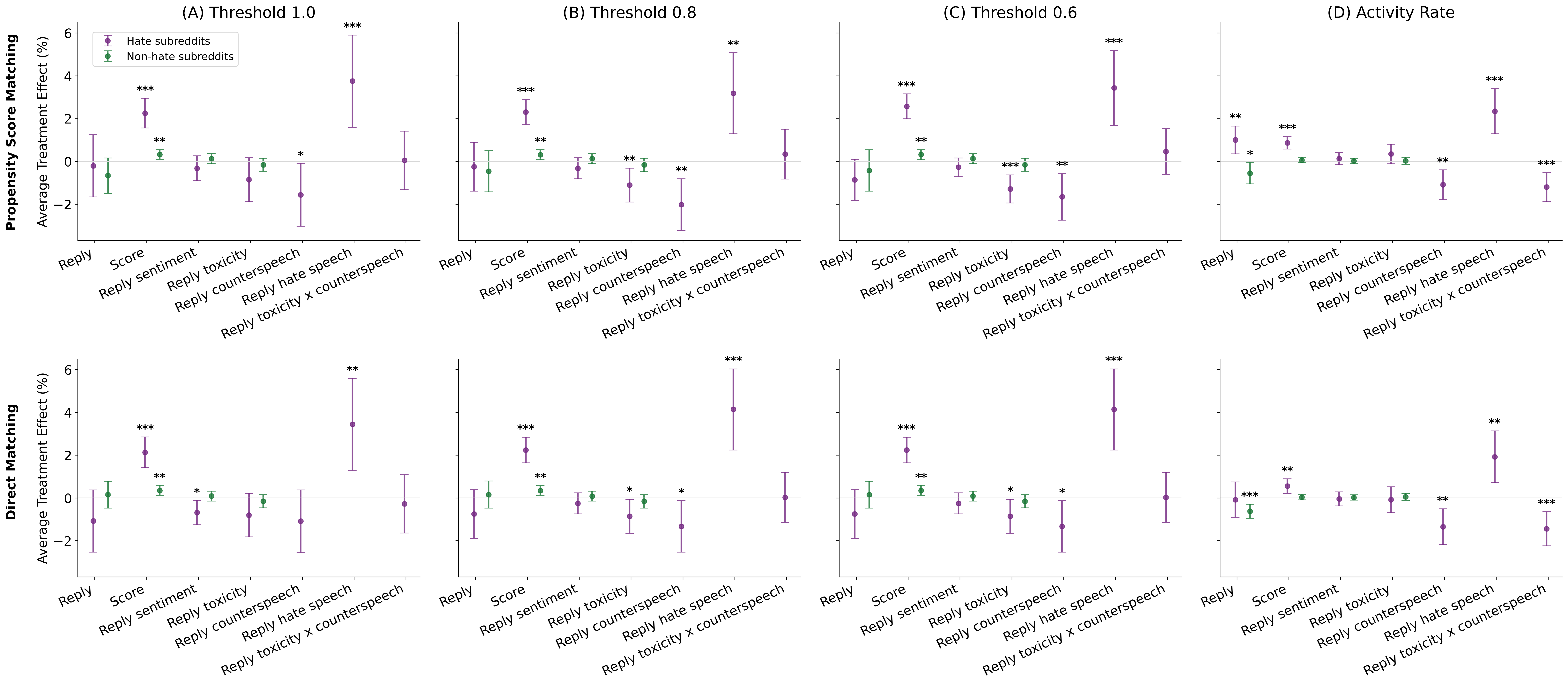}
    \caption{Average treatment effects of counterspeech replies, hate speech replies, and other key variables, with different variations in study design. Each row of subplots shows a different matching strategy. (A, B, C) show the effects when using different thresholds for counterspeech and hate speech detection. (D) shows the results when using a negative binomial regression to estimate the number of unique threads a newcomer will comment under in the month following their initial comment.}
    \label{fig:regression_robustness_checks}
\end{figure*}

\subsubsection{Model Confidence Thresholds}
In the main text, we report estimates of the effects of counterspeech when using only the predictions where the LLM responded with the same label for counterspeech or hate speech at least four out of the five times it was prompted. While we can be confident the model correctly identified most comments and their replies as hate speech or counterspeech, this approach is not as precise as a more restrictive threshold, and yields a much smaller sample size than a less restrictive, yet still precise, threshold. To check how results vary as a function of our choice in prediction thresholds, we repeated our analysis by changing the counterspeech and hate speech thresholds to 1.0 (Fig.~\ref{fig:regression_robustness_checks}A) and 0.6 (Fig.~\ref{fig:regression_robustness_checks}C). When changing classification thresholds, we find the same overall results, with narrower confidence intervals for the effects of counterspeech. However, we find that the overall effect of toxicity on engagement is negative for looser thresholds but not significant for the most strict threshold.

\subsubsection{Matching Type}
While we report results using propensity score matching in the main text, direct matching performs almost as well (Figure~\ref{fig:covariate_balance}). We report the average treatment effects under this alternative matching method for all thresholds. The results of this approach are virtually identical to the results in the main text, with the exception of the most restrictive counterspeech threshold, where the effects of counterspeech are no longer significant.

\subsubsection{Alternative Outcome: Activity Rates}
In the main text, we report the effect of counterspeech on a binary outcome of whether or not the newcomer receiving the reply will continue posting in a given hate subreddit. However, an equally pertinent question might be whether counterspeech replies reduce newcomers' overall activity levels in the subreddit. To measure this, we counted the number of unique threads each newcomer posts or comments under in the subreddit during the four week period following their initial comment. Because the outcome is no longer binary, we estimated the effects of counterspeech on newcomer activity rates using a negative binomial regression model. The average treatment effects are shown in Figure~\ref{fig:regression_robustness_checks}D. Again, the effects of counterspeech and hate speech replies are similar to what is shown in the main text. However, this model suggests that elevated toxicity in counterspeech can play a role in making newcomers less active in their given hate community.

\subsubsection{Subreddit Size Thresholds}

\begin{figure}[h]
    \centering
    \includegraphics[width=0.9\columnwidth]{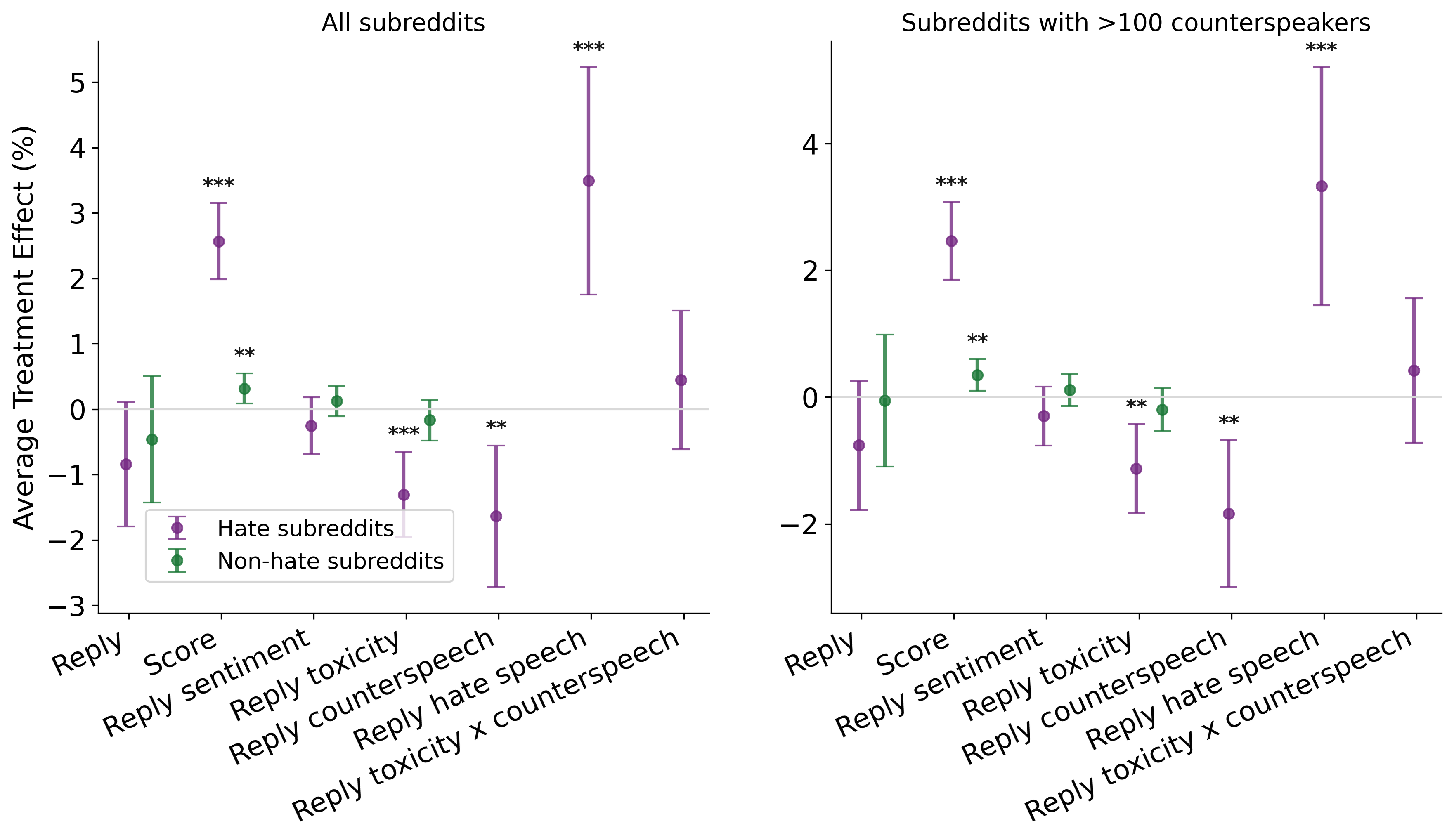}
    \caption{Average treatment effects of counterspeech replies, hate speech replies, and other key variables, with different inclusion criteria for subreddits in the study. The left subplot shows results when including users from all subreddits, while the right subplot shows results when including users from subreddits with at least 100 counterspeech replies to newcomers.}
    \label{fig:robustness_sample_size}
\end{figure}

Given that the subreddits in our study vary greatly in size, we had to make decisions about whether to exclude subreddits that were too small from the analysis, since they could add noise, and the quality of matches in those subreddits may not be as good. However, there was no clear threshold to determine whether to include or exclude a subreddit from our study. In the main text, we report results for all subreddits with at least 10 counterspeakers replying to newcomers, a rather relaxed criterion. Figure~\ref{fig:robustness_sample_size} shows results when including all subreddits in the analysis, as well as the results when only including subreddits with at least 100 counterspeakers replying to newcomers. In both cases, the results are nearly identical to those reported in the main text.

\subsubsection{Analysis of Later Posts}

\begin{figure}
    \centering
    \includegraphics[width=0.9\columnwidth]{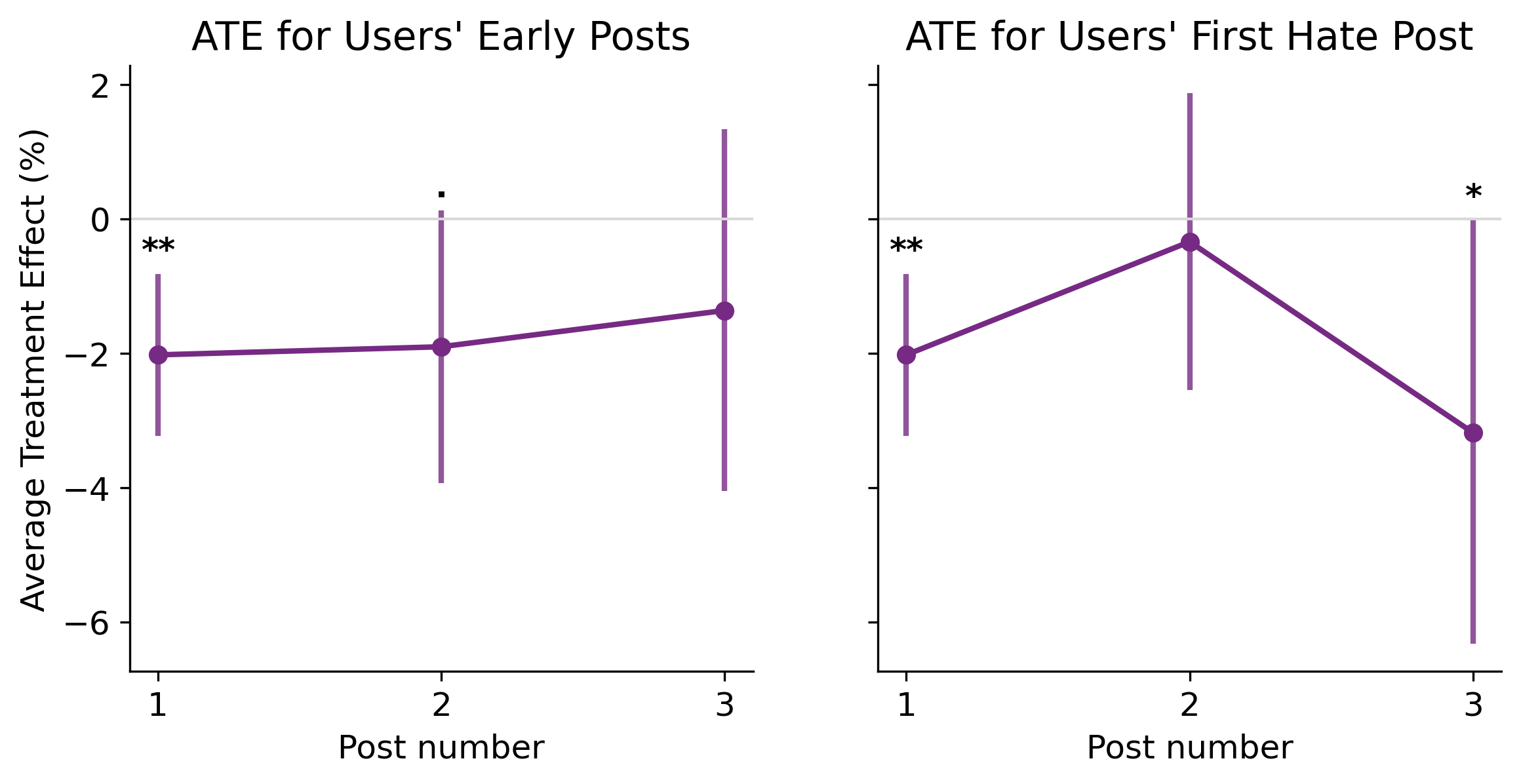}
    \caption{Average treatment effects of counterspeech on user retention for replies to users' first, second, and third comments in hate subreddits. The left subplot shows treatment effects for all users who used hate speech (regardless of if they posted comments containing hate speech earlier), while the subplot on the right excludes users from the sample if they have posted hate speech in the subreddit before.}
    \label{fig:ate_later_posts}
\end{figure}

In addition to studying the effects of counterspeech on users making their first comment in a given hate subreddit, we analyze the effects of counterspeech on users making their second or third comments. Figure~\ref{fig:ate_later_posts} shows the average treatment effects of counterspeech on these comments, revealing that the effect of counterspeech is not significant for users who post a second or third time (except for users commenting for the third time, given that their prior comments did not contain hate speech). However, the magnitude of all the effects are still negative. Importantly, the sample size of users diminishes when analyzing later comments (the sample size for second comments is approximately one half that of first comments, and the sample size for third comments is one third that of first comments), meaning that the lack of significance we observe for later comments could simply be due to smaller sample sizes. Alternatively, users who make multiple posts may also be more resistant to counterspeech, though since we only analyze second and third comments, this represents a rather shallow comparison of newcomers to more established hate users. It is also important to note that even users who comment a second or third time have a much higher baseline probability of continuing to post in a given hate subreddit than newcomers (Fig.~\ref{fig:prob_continue}), even if the magnitudes of relative effects are similar. We leave more in-depth analyses of the effects of counterspeech on highly active members of hate subreddits to future work.

\subsubsection{Summary of Robustness Checks}

The robustness checks performed show largely consistent results regarding the effects of counterspeech and hate speech replies. However, the effects of toxicity are more mixed, with some configurations showing that increased toxicity in any reply has negative effects on engagement (Figure~\ref{fig:regression_robustness_checks}B, C), and another showing that increased toxicity in counterspeech has negative effects on overall activity rates (Figure~\ref{fig:regression_robustness_checks}D). Overall, these results strengthen our assertion that counterspeech in general has negative effects on engagement, while adding to the uncertainty of whether toxic counterspeech reduces engagement in hate subreddits.

% LaTeX table generated from CSV
\begin{table*}[ht]
\centering
\small
\setlength{\tabcolsep}{4pt} % reduce padding slightly
\renewcommand{\arraystretch}{1.2}

\begin{tabularx}{\linewidth}{X X p{0.12\linewidth}}
\toprule
\textbf{Newcomer} & \textbf{Reply} & \textbf{Predicted}\\
& & \textbf{Reply Type} \\
\midrule
\rowcolor{gray!10}
Feminism gets a bad rep because it has no place in the western world. & -- & No reply \\
\rowcolor{gray!10}
Tiktok is made by Chinese genius capitalists marketing towards misandristic **** and social justice warriors, not surprised he was banned. Men no longer have a place in social media. & Absolutely agree & Hate speech \\
\rowcolor{gray!10}
This only makes more confidant in my theory that woman only think on themselves and in the fact that they are woman and need special threatment & That's a bit of a generalization, don't you think? & Counterspeech \\
\addlinespace[6pt]
Black for sure, most crimes are commited by blacks & -- & No reply \\
They should stop killing each other then. They're the ones that are 90\% responsible. & More like 99\% and the media highlights the 1\% where they get instant karma & Hate speech \\
Illegals dont belong here & Says who & Counterspeech \\
\addlinespace[6pt]
\rowcolor{gray!10}
Looks like a **** ******. He probably needs to **** to get any ***** & -- & No reply \\
\rowcolor{gray!10}
How do you think he swims so fast. He uses his giant ****** strength. & youre all wrong, he feels no pain from all the weed he smokes & Hate speech \\
\rowcolor{gray!10}
Houstonian here. This apartment complex is in the middle of China Town. I've heard that in Chinese they call black people "Black Ghosts". I wouldn't want ghosts hanging around my apartment complex either. & Thank you I have never heard this before and I will not be referring to them as that. & Counterspeech \\
\addlinespace[6pt]
Saying "educated by tumblr" is the same as saying "educated by dozens of admitted mentally ill radical man-hating feminists". Does that reeaaally sound impressive? & -- & No reply \\
How did they get married? Isn't she supposed to hate white men? Also, isn't a wedding ring literally rape? & Fiancé - they're not married yet. There's still hope for him to escape. & Hate speech \\
I'm divided. I hate feminists but at the same time I think it's ****** for trans people to force other people to call them 'women' when they are not & You're a dipshit. Of course they are. Anyone who says they're a woman is a woman. & Counterspeech \\
\addlinespace[6pt]
\rowcolor{gray!10}
yeah he seems like a sissy people pleaser with no spine. & -- & No reply \\
\rowcolor{gray!10}
They banned me for saying she's on her period, like that's so unheard of, that women get grumpy when they're on the rag. I actually did the math and she really does go on a mad banning/censoring spree once a month for 5 days. She should just admit it. & "I actually did the math and she really does go on a mad banning/censoring spree once a month for 5 days. She should just admit it." Show us!!! & Other speech \\
\rowcolor{gray!10}
Lmao since when does being gay meant that you have to support gay marriage. 90\% of those in a gay marriage look like femme ******* and your gay parades are fucking sexualized and obnoxious degenerate shit. Fuck your ******** doxxing crap. & Yeah but they're FUN! You straight people wish you could have so much fun! & Counterspeech \\
\addlinespace[6pt]
Black men are peaceful? They are the 6\% of population responsible for more than 50\% of violent crime. & -- & No reply \\
Hes not wrong though, Islam is basically fascism disguised as a religion. inb4 "No, its the rElIgIoN oF pEaCe" & You’re only supposed to capitalize the first letter of a sentence & Other speech \\
is it cuz she's a woman and shouldn't be allowed to vote & ok incel & Counterspeech \\
\bottomrule
\end{tabularx}
\caption{Examples of matched newcomer triplets. The predicted reply type is labeled as hate speech/counterspeech if at least 3/5 LLM responses labeled it as such. Shading alternates by matched set.}
\label{tab:triplet_examples}
\end{table*}

\end{document}